\documentclass[runningheads]{llncs}
\usepackage{graphicx}
\usepackage[utf8]{inputenc}
\usepackage[english]{babel}
\usepackage[utf8]{inputenc}
\usepackage{amsmath}
\usepackage{amsfonts}
\usepackage{mathtools}
\usepackage{xcolor}
\usepackage{listings}
\usepackage{ragged2e}
\usepackage{fancyvrb}
\usepackage{fvextra}
\usepackage{verbatimbox}
\usepackage{hyperref}
\usepackage{nicefrac}

\usepackage{comment}

\usepackage{xspace}

\usepackage[shortlabels]{enumitem}
\setlist[enumerate]{nosep,noitemsep,partopsep=0pt,topsep=0pt,parsep=0pt}

\usepackage{color,soul}

\usepackage{enumitem}
\usepackage{array}
\usepackage{hyperref}

\newtheorem{observation}{Observation}

\newcommand{\calN}{\mathcal{N}}
\newcommand{\calA}{\mathcal{A}}

\newcommand{\calS}{\mathcal{S}}
\newcommand{\calL}{\mathcal{L}}

\newcommand{\calM}{\mathcal{M}}
\newcommand{\calT}{\mathcal{T}}
\newcommand{\calE}{\mathcal{E}}
\newcommand{\calB}{\mathcal{B}}
\newcommand{\calC}{\mathcal{C}}
\newcommand{\calX}{\mathcal{X}}
\newcommand{\calF}{\mathcal{F}}
\newcommand{\calD}{\mathcal{D}}
\newcommand{\calG}{\mathcal{G}}

\newcommand{\NN}{\mathbb{N}}

\newcommand{\ia}{\textit{i}}
\newcommand{\ib}{\textit{ii}}
\newcommand{\ic}{\textit{iii}}

\newcommand{\PP}{P}

\newcommand{\udi}[1]{\textcolor{blue}{Udi says: #1}}

\setlist[itemize]{leftmargin=*}
\setlist[enumerate]{leftmargin=*}

\newcommand{\temph}[1]{\textbf{#1}}

\newcommand{\AD}{All-to-All Block Dissemination\xspace}

\newcommand{\LCC}{Longest-Chain Consensus\xspace}

\newcommand{\GCD}{Grassroots Coin Dissemination\xspace}

\newcommand{\calGCD}{\calG\calC\calD}

\newcommand{\calCGD}{\calC\calG\calD}

\begin{document}
\title{Multiagent Transition Systems for\\ Composing Fault-Resilient Protocol Stacks}
   
\titlerunning{Multiagent Transition Systems}
\author{Ehud Shapiro}
\authorrunning{Ehud Shapiro}
\institute{Weizmann Institute of Science, Rehovot, Israel \\
\email{ehud.shapiro@weizmann.ac.il}
}
\maketitle              % typeset the header of the contribution
\begin{abstract}
We present a novel mathematical framework for the specification and analysis of fault-resilient distributed protocols and their implementations, with the following components:
\begin{enumerate}[partopsep=0pt,topsep=0pt,parsep=0pt]
    \item Transition systems that allow the specification and analysis of computations with safety and liveness faults and their fault resilience.
    \item Notions of safe, live and complete implementations among transition systems and their composition, with which the correctness (safety and liveness) and completeness of a protocol stack as a whole follows from each protocol implementing correctly and completely the protocol above it in the stack.
    \item Applying the notion of monotonicity, pertinent to histories of distributed computing systems, to ease the specification and proof of correctness of implementations among distributed computing systems.
     \item Multiagent transition systems, further characterized as centralized/distributed and synchronous/asynchronous; safety and liveness fault-resilience of implementations among them and their composition.
\end{enumerate}
The framework is being employed in the specification of a grassroots ordering consensus protocol stack, with a grassroots dissemination protocol and its implementation of grassroots social networking~\cite{shapiro2023grassroots} and of sovereign cryptocurrencies~\cite{shapiro2022sovereign},  and an efficient Byzantine atomic broadcast protocols~\cite{keidar2022cordial} as initial applications.

\end{abstract}
%
%
%

%% For double-blind review submission, w/o CCS and ACM Reference (max submission space)

%% Keywords
%% comma separated list
  %% \keywords are mandatory in final camera-ready submission

%% \maketitle
%% Note: \maketitle command must come after title commands, author
%% commands, abstract environment, Computing Classification System
%% environment and commands, and keywords command.

\section{Introduction and Related Work}

This paper  presents a mathematical framework for specifying and proving in a compositional way the correctness and fault-resilience of a distributed protocol stack.  Different aspects of this problem have been addressed for almost half a century.

Process calculi have been proposed for the compositional specification and proof of concurrent systems~\cite{hoare1978communicating,milner1980calculus,milner1999communicating}, mostly focusing on synchronous communication, although variants for asynchronous distributed computing have been investigated~\cite{boudol1992asynchrony,fournet1996reflexive}, including their resilience to fail-stop failures~\cite{francalanza2007theory}

Transition systems are a standard way of specifying computing systems without committing to a specific syntax.  The use of transition systems for the specification of concurrent and distributed systems has been investigated extensively~\cite{hesselink1988deadlock,abadi1991existence,lynch1988introduction}, including the notion of implementations among transition systems and their composition~\cite{abadi1993composing,lynch1988introduction,hur2012marriage}.  The composition of implementations has been investigated in the context of multi-phase compilation~\cite{leroy2009formally,paraskevopoulou2021compositional}, where the correctness of the compiler as a whole following from the correctness of each phase in the compilation. Due to the deterministic and centralized nature of compilation, this task did not require addressing questions of liveness, completeness, and fault tolerance.  Transition systems have been also employed to specify and prove the fault-resilience of distributed systems~\cite{wilcox2015verdi}.

Fault-resilient distributed computing, especially the problems of Byzantine Agreement~\cite{shostak1982byzantine}, Byzantine Reliable Broadcast~\cite{bracha1987asynchronous,guerraoui2019consensus,das2021asynchronous}, Byzantine Atomic Broadcast (ordering consensus)\cite{yin2019hotstuff,keidar2021need,giridharan2022bullshark}, and blockchain consensus~\cite{bitcoin}, have been investigated extensively. Methods for reasoning about distributed systems have been developed~\cite{krogh2020aneris,sergey2017programming,lesani2016chapar}, including their fault resilience~\cite{wilcox2015verdi}, and formal frameworks for the specification and proof of distributed systems were developed~\cite{lynch1988introduction,lamport1999specifying,wilcox2015verdi}. However, the reality is that  novel protocols and their proofs, e.g. ~\cite{cristian1995atomic,yin2019hotstuff,keidar2021need,giridharan2022bullshark,Das2021ADD}, are typically presented outside any formal framework, probably due to the sheer complexity of the protocols and their proofs.

To the best of our knowledge, a mathematical framework for specifying and proving in a compositional way the correctness and fault-resilience of distributed protocol stack, in which each protocol implements the protocol above it and serves as a specification for the protocol below it, is novel.  We developed the framework with the goal of specifying and proving the correctness and fault-resilience of a particular protocol stack:  One that  commences with an open dissemination protocol that can support the grassroots formation of a peer-to-peer social network; continues with a protocol for equivocation exclusion that can support sovereign cryptocurrencies and an equivocation-resilient NFT trade protocol~\cite{shapiro2022sovereign}; and culminates in a group consensus protocol for ordering transactions despite Byzantine faults,  namely Byzantine Atomic Broadcast~\cite{keidar2022cordial}.

\begin{figure}[ht]
\centering
\includegraphics[width=13cm]{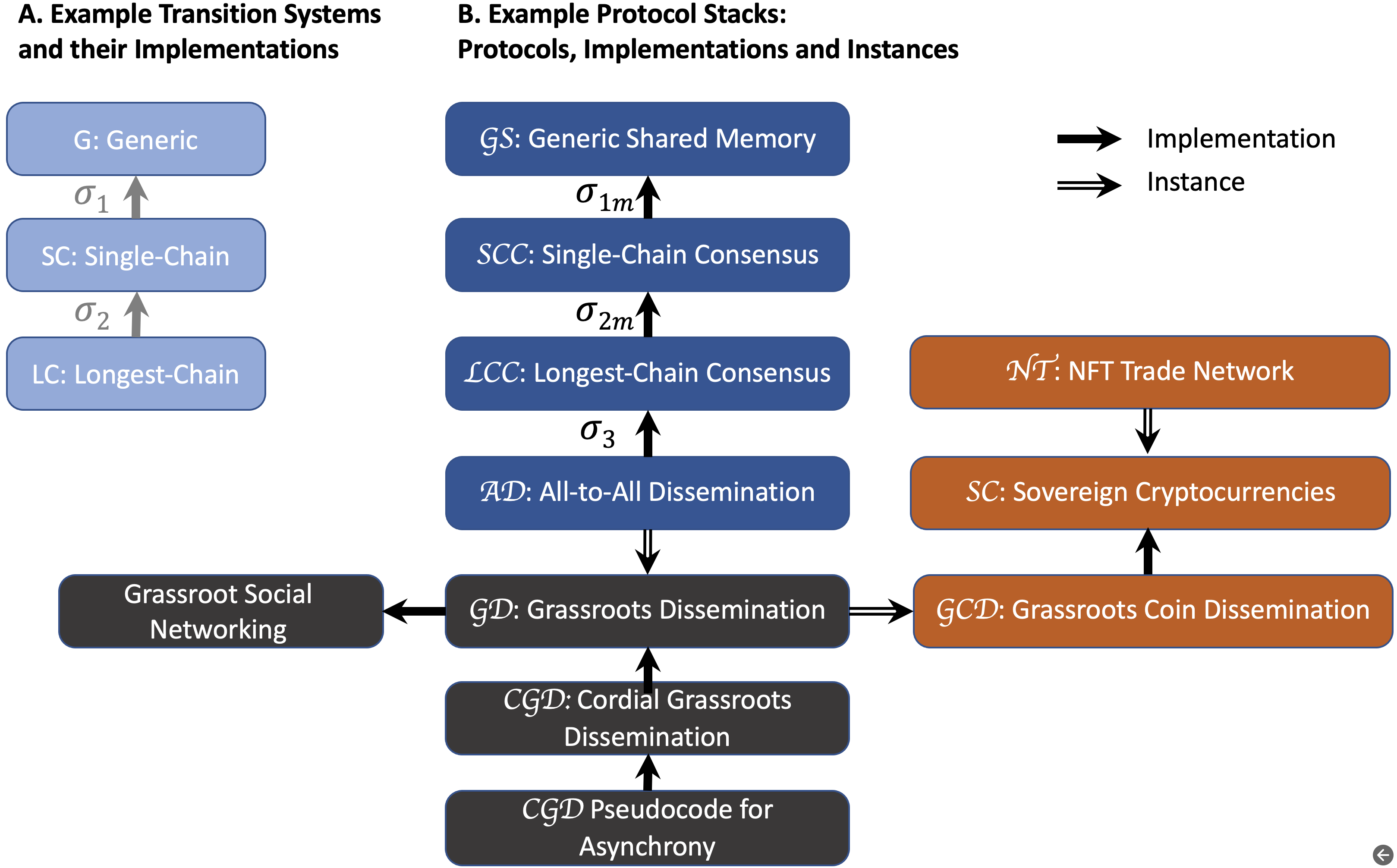}
\caption{Protocol Stacks, Implementations and Instances.
\textbf{A.} Generic (Example \ref{example:G}), Single-Chain (Ex. \ref{example:SC}), and Longest Chain (Ex. \ref{example:LC}) transition systems and their implementations $\sigma_1$ (Proposition \ref{proposition:SC-implements-G}) and $\sigma_2$ (Def. \ref{definition:sigma2} and Prop. \ref{proposition:sigma2-op}).  \textbf{B.} A Protocol stack (in blue), including the Generic Shared-Memory (Ex. \ref{example:GS}), Single-Chain Consensus (Ex. \ref{example:SCC}), \LCC (Ex. \ref{example:LCC}), and \AD (Ex. \ref{example:AD}) protocols and their implementations $\sigma_{1m}$ (Prop. \ref{proposition:SCC-implements-GS}), $\sigma_{2m}$ (Prop. \ref{proposition:LCC-implements-SCC}), and $\sigma_3$ (Prop. \ref{proposition:AD-can-implement-LCC}). Results from companion papers that apply the mathematical framework presented here are in grey~\cite{shapiro2023grassroots} and light brown~\cite{shapiro2022sovereign}: The Grassroots Dissemination protocol $\calG\calD$, an instance of $\calA\calD$, implementing grassroots social networking and implemented by the Cordial Grassroots Dissemination protocol $\calCGD$, which in turn has a pseudocode implementation for the model of asynchrony (not a multiagent transition system), demonstrating the feasibility of a realistic (TCP-based) implementation of the protocol and the protocol stack it supports. The \GCD protocol $\calGCD$, an instance of $\calG\calD$,  and its implementation of the Sovereign Cryptocurrencies protocol $\calS\calC$, which in turn is an instance of the NFT Trade Protocol $\calN\calT$}.

\label{figure:example-ts}
\end{figure}

Here, we present, prove correct, and analyze the fault-resilience of two abstract protocol stacks, depicted in Figure \ref{figure:example-ts}, as example applications of the mathematical framework.  A more concrete, complex, and practical protocol stack based on the blocklace (a partially-ordered generalization of the blockchain) is presented and analyzed  elsewhere~\cite{shapiro2022blocklace}, using the mathematical framework developed here.

A key objective of this work is the development grassroots protocols that can be deployed independently at different locations and over times,  initially with disjoint communities operating the protocol independently, and over time---once connected--forming an ever-growing interacting networked community. Here we characterize the notion of a protocol being grassroots algebraically and operationally,  analyze whether protocols in the abstract protocol stack are grassroots, and discuss whether client-server protocols (e.g., all major digital platforms), consensus protocols (e.g. reliable broadcast, Byzantine agreement), majoritarian decision making protocols (e.g. democratic voting),  and protocols that employ a non-composable data structure (e.g., blockchain), are grassroots, and if not, then whether and how can they be made so.

Our approach is different from that of universal composability~\cite{canetti2001universally}, devised for the analysis of cryptographic protocols, in at least two respects:  First, it does not assume, from the outset, a specific notion of communication.  Second, its notion of composition is different: Universal composability uses function composition as is common in the practice of protocol design (e.g.~\cite{keidar2021need,das2021asynchronous,princehouse2014mica}). Here,  we do not compose protocols, but compose implementations among protocols, resulting in a new \underline{single} implementation that realizes the high-level protocol using the primitives of the low-level protocol. For example, it seems that the universality results of Sections \ref{section:ts} and \ref{subsection:mts}  cannot be expressed in the model of universal composability.

In the rest of the paper Section \ref{section:ts} presents transition systems,  implementations among them, and the composition of such implementations, and includes the example protocol stack  of Figure \ref{figure:example-ts}A.  It also introduces the notion of monotonicity of transition systems~\cite{pilkiewicz2011essence,hawblitzel2017ironfleet}, and shows that it can ease the proof of correctness of an implementation.
Section \ref{subsection:mts} presents multiagent transition systems, further characterized as centralized or distributed, with the latter being synchronous or asynchronous, and includes the example multiagent protocol stack of Figure \ref{figure:example-ts}B. 
Section \ref{subsection:mtsf} introduces safety faults and liveness faults,   implementations that are resilient to such faults, and their composition.
Section \ref{subsection:protocol} introduces formally the notion of a protocol as a family of multiagent transition systems and provides an example.
Section \ref{section:conclusions} concludes the paper.
Proof are relegated to Appendix \ref{appendix-section:proof}.

\section{Transition Systems, Implementations and their Composition}\label{section:ts}

Here, we introduce the notions of transition systems, implementations among them, and their composition, together with the examples of Figure \ref{figure:example-ts}A.

\subsection{Transition Systems and Their Implementation}\label{subsection:tsfi}

Given a set $S$, $S^*$ denotes the set of sequences over $S$, $S^+$ the set of nonempty sequences over $S$, and $\Lambda$ the empty sequence.  Given $x, y \in S^*$,  $x\cdot y$  denotes the concatenation of $x$ and $y$, and  $x \preceq y$  denotes that $x$ is a prefix of $y$. 
Two sequences $x, y \in S^*$ are \emph{consistent} if $x \preceq y$ or $y \preceq x$, \emph{inconsistent} otherwise.

\begin{definition}[Transition System, Computation, Run]\label{definition:ts}
Given a set $S$, referred to as  \temph{states},  the \temph{transitions} over $S$ are all pairs $(s,s') \in S^2$, also written $s \rightarrow s'$.
A \temph{transition system} $TS=(S,s_0,T,\lambda)$ consists of a set of states $S$, an initial state $s_0 \in S$, a set of  \temph{correct transitions} $T \subseteq S^2$, and a \temph{liveness condition} $\lambda$ which is a set of sets of correct transitions; when $\lambda$ is omitted the default liveness condition is $\lambda = \{T\}$.
%The set $s \rightarrow * = \{ \hat s~ |~ s \xrightarrow{} \hat s \in T\}$ is the \temph{outgoing transitions} of $s$.   
A \temph{computation} of $TS$ is a sequence of transitions $r= s \xrightarrow{} s' \xrightarrow{}  \cdots \subseteq S^2$.   A \temph{run} of $TS$ is a computation that starts from $s_0$. 
\end{definition}

Recall that safety requires that bad things don't happen, and liveness that good things do happen, eventually. For example, ``a transition that is enabled infinitely often is eventually taken''. Heraclitus said that you cannot step into the same river twice.  Similarly, in a transition system you cannot take the same transition in different states as, by definition, it is a different transition. Hence, a liveness condition is a requirements on sets of transitions,  rather than on individual transitions. For example, the set can be all transitions in which `$p$ receives message $m$ from $q$', even if the local state of $p$ or of other agents differ.  In multiagent transition systems, defined below, liveness may require each agent to act every so often.  To specify such a liveness condition, all transitions by the same agent would form a set.  

\begin{definition}[Safe, Live and Correct Run]\label{definition:ts-slc}
Given a transition system  $TS=(S,s_0,T,\lambda)$, a computation $r$ is \temph{safe}, also $r \subseteq T$, if every transition of $r$ is correct, and  $s \xrightarrow{*} s' \subseteq T$ denotes the existence of a safe computation (empty if $s=s'$) from $s$ to $s'$. 

A transition $s'\rightarrow s'' \in S^2$ is \temph{enabled on $s$} if $s=s'$.
A run is \temph{live wrt $L \in \lambda$} if either $r$ has a nonempty suffix in which no transition in $L$ is enabled, or every suffix of $r$ includes an $L$ transition. A run $r$ is \temph{live} if it is live wrt every $L \in \lambda$.
A run $r$ is \temph{correct} if it is safe and live.
\end{definition}

\begin{observation}[Final State]\label{observation:final-state}
A state is \temph{final} if no correct transition is enabled on it. 
A live computation is finite only if its last state is final.
\end{observation}

The following is an example of a generic transition system over a given set of states.
Here and in the other examples in this section the liveness condition $\lambda$ is omitted and a computation is live if it is live wrt the correct transitions.

\begin{example}[G: Generic]\label{example:G}
Given a set of states $S$ with a designated initial state $s0 \in S$, a generic transition system  over $S$
is $G =(S,s0,TG)$ for some $TG\subseteq S^2$.
\qed\end{example}

\begin{definition}[Specification; Safe, Live, Correct and Complete Implementation]\label{definition:implementation}
Given two transition systems $TS=(S,s_0,T,\lambda)$ (the \temph{specification}) and $TS'=(S',s'_0,T',\lambda')$, \temph{an implementation of $TS$ by $TS'$} is a function $\sigma : S' \rightarrow S$ where $\sigma(s'_0) = s_0$, in which case the pair $(TS', \sigma)$ is referred to as  \temph{an implementation of $TS$}. Given a computation $r'= s'_1\rightarrow s'_2 \rightarrow \ldots$ of $TS'$, $\sigma(r')$ is the (possibly empty) computation $\sigma(s'_1) \rightarrow \sigma(s'_2) \rightarrow \ldots$, obtained from the $\sigma(s'_1)\rightarrow \sigma(s'_2) \rightarrow \ldots$ by removing consecutively repetitive elements so that $\sigma(r')$ has no \temph{stutter transitions} of the form $s \rightarrow s$. The implementation $(TS', \sigma)$ of $TS$ is \temph{safe/live/correct} if $\sigma$ maps every safe/live/correct $TS'$ run $r'$ to a safe/live/correct $TS$ run $\sigma(r')$, respectively, 
and is \temph{complete} if every correct run $r$ of $TS$ has a correct run $r'$ of $TS'$ such that $\sigma(r')=r$.
\end{definition}

\begin{definition}[$\sigma$: Locally Safe, Productive, Locally Complete]
Given two transition systems $TS=(S,s_0,T,\lambda)$ and $TS'=(S',s'_0,T',\lambda')$ and an implementation $\sigma: S' \mapsto S$. 
Then $\sigma$ is:
\begin{enumerate}[partopsep=0pt,topsep=0pt,parsep=0pt]
    \item \temph{Locally Safe} if $s'_0 \xrightarrow{*} x'_1 \xrightarrow{} x'_2 \subseteq T'$ implies that  $s_0 \xrightarrow{*} x_1  \xrightarrow{*} x_2 \subseteq T$ for $x_1 = \sigma(x'_1)$ and $x_2= \sigma(x'_2)$ in $S$. If $x_1 = x_2$ then the $T'$ transition  $x'_1 \xrightarrow{} x'_2$ \temph{stutters} $T$.
    
     \item \temph{Productive} if for every $L \in \lambda$ and every correct run $r'$ of $TS'$,
     either $r'$ has a nonempty suffix $r''$ such that $L$ is not enabled in $\sigma(r'')$,
     or every suffix $r''$ of $r'$  \temph{activates} $L$, namely $\sigma(r'')$ has an $L$-transition.
    
    \item \temph{Locally Complete} if  $s_0 \xrightarrow{*} x_1\xrightarrow{} x_2 \subseteq T$, implies that
    $s'_0 \xrightarrow{*} x'_1 \xrightarrow{*} x'_2 \subseteq T'$ for some $x'_1, x'_2 \in S'$ such that $x_1= \sigma(x'_1)$ and
     $x_2 = \sigma(x'_2)$. 
\end{enumerate}
\end{definition}

\begin{proposition}[$\sigma$ Correct]\label{proposition:sigma-correct}
If an implementation $\sigma$ is locally safe and productive then it is correct, and if in addition it is locally complete then it is complete.  
\end{proposition}

Intuitively, in an implementation $(TS',\sigma)$ of $TS$, $TS'$ can be thought of as the `virtual hardware' (e.g. the instruction set of a virtual machine or the machine language of an actual machine) and $\sigma$ as specifying a `compiler', that compiles programs in the high-level language $TS$ into machine-language programs in $TS'$.  The mapping $\sigma$ from $TS'$ to $TS$ is in inverse direction to that of a compiler; it thus specifies the intended behavior of compiled programs in terms of the behavior of their source programs, and in doing so can serve as the basis for proving a compiler correct. Note, though, that transition systems have no formal syntax, and can be thought of as specifying the operational semantics of existing or hypothetical programming languages.

Preparing an example implementation,  we present the universal single-chain transition system SC, and then show how it can implement any generic transition system G, justifying the title `universal'.

\begin{example}[SC: Single-Chain]\label{example:SC}
Given a set $S$ with a designated initial state $s0\in S$, the single-chain transition system over $S$ is SC $=(S^+,s0,TSC)$, where $TSC$ includes every transition $x \rightarrow x\cdot s$ for every $x \in S^*$ and $s \in S$.
\qed\end{example}
Namely, an SC run can generate any sequence over $S$.

From a programming-language perspective, some transition systems we will be concerned with are best viewed as providing the operational semantics for a set of programs over a given domain. 
With this view, in the current abstract setting, the programming of a transition system, namely choosing a program from this potentially-infinite set of programs, is akin to identifying a (computable) subset of the transition system.  In our example, for the universal single-chain transition system SC to implement a specific instance of the generic transition system G, an instance of SC has to be identified that corresponds to the transitions of G, as shown next.
But first we define the notion of a transition system subset.

\begin{definition}[Transition System Subset]\label{definition:instance}
Given a transition system $TS=(S,s_0,T,\lambda)$,  a transition system $TS'=(S',s'_0,T',\lambda')$ is an \temph{instance} (or subset) of $TS$, $TS' \subseteq TS$, if
$s'_0 = s_0$, $S'\subseteq S$, $T' \subseteq T$, and $\lambda'$ is $\lambda$ restricted to $T'$,  $\lambda' := \{ L \cap T' |  L\in \lambda\}$.
%and $TS'$ is \temph{relatively complete} wrt $T$: $s_0 \xrightarrow{*} s \subseteq T$ with all states in $S'$ implies that  $s_0 \xrightarrow{*} s \subseteq T'$.
\end{definition}
The definition suggests at least two specific ways to construct an instance:  Choosing a subset of the states and restricting the transitions to be only among these states; or choosing a subset of the transitions.  Specifically, (\ia)  Choose some $S' \subset S$ and define  $T':=T/S'$, namely $T' := \{(s\xrightarrow{}s' \in T ~:~ s, s' \in S'\}$. (\ib) Choose some $T' \subset T$.  We note that in practice there must be restrictions on the choice of a subset; to begin with, $S'$ and $T'$ should be computable.

We want to show that the universal single-chain transition system can implement any generic transition system.  Hence the following definition:

\begin{definition}[Can Implement]\label{definition:can-implement}
Given transition systems $TS=(S,s_0,T,\lambda)$, $TS'=(S',s_0',T',\lambda')$, $TS'$ \temph{can implement $TS$}  if there is an instance $TS''=(S'',s_0',T'',\lambda'')$, $TS''  \subseteq TS'$ and a correct and complete implementation $\sigma: S'' \mapsto S$ of $TS$ by $TS''$.
\end{definition}

Next we  demonstrates the application of the definitions above:
\begin{proposition}\label{proposition:SC-implements-G}
The single-chain transition system SC over $S$ can implement any generic transition system 
G over $S$.
\end{proposition}

\subsection{Composing Implementations}

The key property of correct and complete implementations is their transitivity:
\begin{proposition}[Transitivity of Correct \& Complete Implementations]\label{proposition:impelementation-transitivity}
The composition of safe/live/correct/complete   implementations is safe/live/correct/complete, respectively.
\end{proposition}

%The proof in the Appendix includes explanatory Figure \ref{figure:proposition-1}.

Our next example is the longest-chain transition system, which can be viewed as an abstraction of the longest-chain consensus protocols (e.g. Nakamoto~\cite{bitcoin}), since its consistency requirement entails that only the longest chain may be freely extended; other chains are bound to copy their next sequence element from a longer chain till they catch up, if ever, and only then may contribute a new element to the chain.

\begin{example}[LC: Longest-Chain]\label{example:LC}
Given a set $S$ and $n> 0$,  the LC longest-chain transition system over $S$, LC $~= ((S^*)^n,c0,TLC)$, has sets of $n$ sequences over $S$ as states, referred to as $n$-chain configurations over $S$, initial state $c0 = \Lambda^n$, and as transitions $TLC$ every $c \rightarrow c'$ where $c'$ is obtained from $c$ by extending one sequence $x \in c$ to $x\cdot s$, $s\in S$, provided that either $x$ is a longest sequence in $c$ or $x\cdot s$ is a prefix of some $y \in c$.
\qed\end{example}

We wish to prove that the longest-chain transition system LC can implement the single-chain transition system SC, and by transitivity of correct implementations, also implement any generic transition systems G.  The mathematical machinery developed next will assist in achieving this.

\subsection{Monotonic Transition Systems for Distributed Computing}\label{subsection:monotonic}

Unlike shared-memory systems,  distributed systems have a state that increases in some natural sense as the computation progresses, e.g. through accumulating messages and extending the history of local states.  This notion of monotonicity, once formalized, allows a simpler and more powerful mathematical treatment of transition systems for distributed computing.

So far we have used $\preceq$ to denote the prefix relation.
In the following we also use $\prec$ to denote a partial order, with $\preceq$ also denoting any non-strict partial order; the intention should be clear from the context.

\begin{definition}[Partial Order]\label{definition-po}
A partial order on a set $S$ is denoted by $\prec_S$ (with $S$ omitted if clear from the context), where $s\preceq s'$ stands for $s \prec s' \vee s =s'$.
The partial order is  \temph{unbounded} if for every $s \in S$ there is an $s' \in S$ such that $s \prec s'$.
We say that $s \ne s' \in S$ are \temph{consistent} wrt $\prec$ if $s \prec s'$ or $s' \prec s$.
\end{definition}

It is often possible to associate a partial order with a distributed system, wrt which the local state of each agent only increases. Therefore we focus on the following type of transition systems:
 
\begin{definition}[Monotonic \& Monotonically-Complete Transition System]\label{definition:monotonic}
Given a partial order $\prec$ on $S$,
a transition system $TS=(S,s_0,T,\lambda)$ is \temph{monotonic} with respect to $\prec$ if $s \rightarrow s' \in T$ implies $s \preceq s'$.
It is \temph{monotonically-complete} wrt $\prec$ if, in addition,  $s_0 \xrightarrow{*} s \subseteq T$ and $s \preceq s'$ implies that $s \xrightarrow{*} s' \subseteq T$.
\end{definition}
Namely, computations of a monotonically-complete transition system not only ascend in the partial order, but may also reach, from any state, any larger state in the partial order.
Note that since the partial order is unbounded, a monotonically-complete transition system has no final states. 
Many applications of this framework, including the examples herein, require proving that a transition system is monotonically-complete.  The following approach is often helpful:

\begin{definition}[$\epsilon$-Monotonic Completeness]\label{definition:epsilon-monotonic}
A transition system $TS=(S,s_0,T,\lambda)$, monotonic wrt a partial order $\prec$ on $S$, 
is \temph{$\epsilon$-monotonically-complete} wrt $\prec$ if 
infinite ascending chains in $\prec$ are unbounded and
$s \prec s''$ implies that there is a transition $s \rightarrow s'  \in T$ such that $s\prec s'$ and $s'\preceq s''$.
\end{definition}

\begin{proposition}[$\epsilon$-Monotonic Completeness]\label{proposition:epsilon-monotonic}
A transition system that is $\epsilon$-monotonically-complete is monotonically-complete.
\end{proposition}
\begin{proof}[of Proposition \ref{proposition:epsilon-monotonic}]
Assume a transition system $TS=(S,s_0,T,\lambda)$ that is $\epsilon$-monotonically-complete wrt a partial order $\prec$ on $S$, and let $s \preceq s''$ for $s, s'' \in S$.  We construct a computation $s \xrightarrow{*}s'' \in T$ iteratively as follows:  Given $(s,s'')$,
if $s=s''$ we are done, else let the next transition of the computation be
$s \rightarrow s' \in T$ for some $s'$ for which $s \prec s'$, $s' \preceq s''$, which exists by assumption, and iterate with $(s', s'')$.
The constructed sequence is an ascending chain bounded by $s''$, which is finite by assumption,  hence the iterative construction terminates with $(s'',s'')$.
\qed \end{proof}

When transition systems are monotonically-complete wrt a partial order, the following Definition \ref{definition:op-implementation} and Theorem \ref{theorem:sigma-op} can be a powerful tool in proving that one can correctly implement the other.

\begin{definition}[Order-Preserving Implementation]\label{definition:op-implementation}
Let transition systems $TS=(S,s_0,T,\lambda)$ and $TS'=(S',s'_0,T',\lambda')$ be monotonic wrt the partial orders $\prec$ and $\prec'$, respectively.
Then an implementation $\sigma : S' \rightarrow S$ of $TS$ by $TS'$ is \temph{order-preserving} wrt  $\prec$ and $\prec'$ if:
\begin{enumerate}[partopsep=0pt,topsep=0pt,parsep=0pt]
    \item \temph{Up condition:} 
     $y_1 \preceq' y_2$ implies that $\sigma(y_1) \preceq \sigma(y_2)$
    \item \temph{Down condition:}  $s_0 \xrightarrow{*}x_1 \subseteq T$, $ x_1 \preceq x_2$ implies that there are $y_1,y_2 \in S'$ such that $x_1= \sigma(y_1)$, $x_2= \sigma(y_2)$, $s'_0 \xrightarrow{* }y_1 \subseteq T'$ and $y_1 \preceq' y_2$.
\end{enumerate}
\end{definition}

Note that if $\preceq'$ is induced by $\sigma$ and $\preceq$, namely defined by $y_1 \preceq' y_2$ \temph{if} $\sigma(y_1) \preceq \sigma(y_2)$, then the Up condition holds trivially. 
%Furthermore, if $\preceq$ is strict, then $y_1 \simeq' y_2$ implies that $\sigma(y_1) = \sigma(y_2)$.  
%In the following we assume that the partial order of the specification is strict; this simplifies the mathematical framework, reduces notational clutter, and is sufficient for our subsequent proofs.
The following Theorem is the linchpin of the proofs of protocol stack theorems here and in other distributed computing applications of the framework.

\begin{theorem}[Correct \& Complete Implementation Among Monotonically-Complete Transition Systems]\label{theorem:sigma-op}
Assume two transition systems $TS=(S,s_0,T,\lambda)$ and $TS'=(S',s'_0,T',\lambda')$, monotonically-complete wrt the unbounded partial orders $\prec$ and $\prec'$, respectively, and an implementation $\sigma : S' \rightarrow S$  of $TS$ by $TS'$. 
%Furthermore, assume that $\preceq$ is %strict and 
%unbounded. 
If $\sigma$ is order-preserving and productive then it is correct and complete.
\end{theorem}

If all transition systems in a protocol stack are  monotonically-complete,   then Theorem \ref{theorem:sigma-op} makes it sufficient to establish that an implementation of one protocol by the next is order-preserving and productive to prove it correct.
A key challenge in showing that Theorem \ref{theorem:sigma-op} applies is proving that the implementation satisfies the Down condition (Def. \ref{definition:op-implementation}), which can be addressed by finding an `inverse' to $\sigma$ as follows:

%This can be achieved by a construction as follows:

\begin{observation}[Representative Implementation State]\label{observation:representatives}
Assume $TS$ and $TS'$ as in Theorem \ref{theorem:sigma-op} and an implementation $\sigma: S' \xrightarrow{} S$ that satisfies the Up condition of Definition \ref{definition:op-implementation}.  If there is a function $\hat{\sigma}: S \xrightarrow{} S'$ such that $x = \sigma(\hat{\sigma}(x))$ for every $x \in S$, and $x_1 \preceq x_2$ implies that $\hat{\sigma}(x_1) \preceq' \hat{\sigma}(x_2)$, then $\sigma$ also satisfies the Down condition.
\end{observation}

\begin{proposition}\label{proposition:LC-implements-SC}
LC can implement SC.
\end{proposition}
\begin{proof}[outline of Proposition \ref{proposition:LC-implements-SC}]
We show that both SC and LC are monotonically-complete wrt the strict prefix relation $\prec$ (Observations \ref{observation:SC-MC}, \ref{observation:LC-MC}) and that the implementation $\sigma_2$ of SC by LC is order preserving and productive (Proposition \ref{proposition:sigma2-op}).
Hence, according to Theorem \ref{theorem:sigma-op}, $\sigma_2$  is correct and complete.
\qed \end{proof}

\begin{definition}[$\sigma_{2}$]\label{definition:sigma2}
The implementation $\sigma_{2}$ maps every $n$-chain configuration $c$ to the longest chain in $c$ if it is unique, and is undefined otherwise.
\end{definition}

In our example, the longest-chain transition system LC implements the single-chain transition system SC.  But SC does not implement the generic transition system G -- an instance of it, SC1, does.  So, in order to prove that LC can implement G, solely based on the implementation of SC by LC, without creating a custom subset of LC for the task, the following Proposition is useful.

\begin{proposition}[Restricting a Correct Implementation to an Instance]\label{proposition:subset}
Let $\sigma: C2 \mapsto S1$ be an order-preserving implementation of $TS1=(S1,s1,T1,\lambda1)$ by $TS2=(C2,s2,T2,\lambda2)$, monotonically-complete respectively with $\prec_1$ and $\prec_2$.  Let 
$TS1'=(S1',s1,T1',\lambda1') \subseteq TS1$ and $TS2'=(C2',s2,T2',\lambda2') \subseteq TS2$ defined by $C2' := \{s \in C2 ~:~ \sigma(s) \in S1'\}$, with $T2' := T2/C2'$, and assume that both instances are also monotonically-complete wrt $\prec_1$ and $\prec_2$, respectively. 
If  $y_1 \xrightarrow{} y_2 \in T2' ~\&~ \sigma(y_1) \in S1'$ implies that $\sigma(y_2) \in S1'$ then the restriction of $\sigma$ to $C2'$ is a correct and complete implementation of $TS1'$ by $TS2'$.
\end{proposition}

\begin{corollary}\label{corollary:LC-universal}
The longest-chain transition system LC is universal for generic transition systems.
\end{corollary}

More generally, Proposition \ref{proposition:subset} is useful in the following scenario.  Assume that protocols are specified via transition systems, as elaborated below.  Then in a protocol stack of, say, three protocols P1, P2, P3, each implementing its predecessor, it may be the case that for the middle protocol P2 to implement the full top protocol P1, an instance P2$'$ of P2 is needed. But, it may be desirable for P3 to implement the full protocol P2, not just its subset P2$'$, as P2 may have additional applications beyond just implementing P1. In particular, there are often application for which an implementation by a middle protocol in the stack is more efficient than an implementation by the full protocol stack. The following proposition enables that, see Figure \ref{figure:subset}. Note that, as shown in the figure, the implementing transition system $TS2$ that implements $TS1$ could in turn be an instance of a broader unnamed transition system.

\section{Multiagent Transition Systems: Centralized, Distributed, Synchronous and Asynchronous}\label{subsection:mts}

\subsection{Multiagent Transition Systems}\label{subsection:mts-ss}

Assume a set $\Pi$ of \emph{agents}.  While the set of all agents $\Pi$ could in principle be infinite (think of all the agents that are yet to be born), when we refer to a particular set of agents $P \subseteq \Pi$ we assume $P$ to be finite.  In the following, $a \ne b \in X$ is a shorthand for $a \in X \wedge b \in X  \wedge a\ne b$.

In the context of multiagent transition systems, the state of the system is referred to as \temph{configuration}, so as not to confuse it with the \temph{local states} of agents in a distributed multiagent transition system, defined next.  

\begin{definition}[Multiagent Transition System]\label{definition:multiagent}
%\begin{definition}[Multiagent Distributed Transition Systems]\label{definition:multiagent}
Given agents $\PP \subseteq \Pi$, a transition system $TS=(C,c_0,T,\lambda)$, with configurations $C$, initial configuration $c0$, correct transitions $T \subseteq C^2$, and a liveness condition $\lambda$ on $T$, is
\temph{multiagent over $P$} if there is a \temph{multiagent partition} $C^2 = \bigcup_{p \in P} CC_p$ of $C^2$ into disjoint sets $CC_p$ indexed by $P$,  $CC_p \cap CC_q = \emptyset$ for every $p\ne q \in P$.  A transition $t= s \rightarrow s' \in CC_p$ is referred to as a \temph{$p$-transition},
 the set of \temph{correct $p$-transitions} $T_p$ is defined by $T_p := T  \cap  CC_p $, for every $p \in P$, and the \temph{multiagent liveness condition} $\lambda$ is a refinement of the multiagent partition of $T$, namely for each $L \in \lambda$, $L \subseteq T_p$ for some $p \in P$.
\end{definition}
Note that $CC_p$ includes all possible behaviors of agent $p$, both correct and faulty, and $T_p$ includes only the agent's correct behaviors.  The multiagent liveness condition considers each agent as autonomous by placing liveness requirements on each agent independently.

\begin{definition}[Safe, Live \& Correct Agents]\label{definition:multiagent-slc}
Given a multiagent transition system $TS=(C,c0,T,\lambda)$ over $P$ and a run $r$ of $TS$, an agent $p$ is \temph{safe in $r$} if $r= c0 \rightarrow c1\rightarrow \ldots$ includes only correct $p$-transitions;  is \temph{live in $r$} if for every $L \in \lambda$ for which $L \subseteq T_p$, $r$ is live wrt $L$; and  is \temph{correct} in $r$ if $p$ is safe and live in $r$. 
%The $p$-run of $r$, $r_p$, is the sequence $c0_p \rightarrow c1_p \ldots$ with stutter transitions removed.
\end{definition}
Note that if $\lambda = \{T_p : p \in P\}$, namely the liveness  condition is the partition of correct transitions to agents, then an agent $p$ is live if it is live wrt its correct $p$-transitions $T_p$.

Next, the generic transition system (Example \ref{example:G}) is modified to be multiagent.
In the generic shared-memory multiagent transition system GS defined next, all agents operate on the same shared global state.  Yet, the transitions of different agents are made disjoint by capturing abstractly the reality of shared-memory multiprocessor systems:  Each configuration incorporates, in addition to a shared global state $s \in S$, also a unique program counter for each agent.  The program counter of agent $p$ is advanced when a $p$-transition is taken.  

\begin{example}[GS: Generic Shared Memory]\label{example:GS}
Given a set of agents $P \subseteq \Pi$ and states $S$ with a designated initial state $s0$, a generic shared-memory multiagent  transition system  over $P$ and $S$,
 $GS =(C,c0,TGS)$, has configurations $C = S \times \calN^P$ that include a shared global state in $S$ and a program counter  $i_p \in \calN$ for each agent $ p \in P$, initial state $c0 = (s0,\{0\}^P)$, and transitions  $TGS = \bigcup_{p \in P} TGS_p \subseteq C^2$, where each $p$-transition $(s,i) \rightarrow (s',i') \in TGS_p$ satisfies $i'_p = i_p+1$ and $i'_q = i_q$ for every $q \ne p \in P$.
\qed\end{example}
Note that $TGS$ is arbitrary, and different agents may or may not be able to change the shared global state in the same way.  But each transition identifies the agent $p$ making the change by advancing $p$'s program counter.

Next, the single-chain transition system SC (Example \ref{example:SC}) is modified to
the multiagent transition system for single-chain  consensus SCC.  As SCC is monotonic, program counters are not needed; it is sufficient to identify the agent contributing the next element to the shared global chain to make transitions by different agents disjoint.

\begin{example}[SCC: Single-Chain  Consensus]\label{example:SCC}
Given a set of agents $P \subseteq \Pi$ and a set $S$, the single-chain consensus multiagent transition system over $P$ and $S$ is SCC $=((S \times P)^*,\Lambda,TSCC)$, 
with each configuration being a sequence of agent-identified states $(s,p)$ of a state $s \in S$ and an agent  $p \in P$, and  $TSCC$ includes every transition $x \rightarrow x\cdot (s,p)$ for every $x \in (S \times P)^*$, $s \in S$ and $p \in P$.
\qed\end{example}
Namely, an SCC run can generate any sequence of agent-identified elements of $S$, where any agent may contribute any element to any position in the sequence.

Next,  we show that SCC can implement GS, making single-chain consensus universal for shared-memory multiagent transition systems.

\begin{proposition}\label{proposition:SCC-implements-GS}
SCC over $P \subseteq\Pi$ and $S$ can implement any generic shared-memory multiagent transition system 
GS over $P$ and $S$.
\end{proposition}

\subsection{Centralized and Distributed Multiagent Transition Systems}\label{subsection:mts-cd}

Having introduced centralized/shared-memory multiagent transition systems, and before introducing  distributed ones, we formalize the two notions:

\begin{definition}[Centralized and Distributed Multiagent Transition System]\label{definition:multiagent-cd}
A multiagent transition system $TS=(C,c_0,T,\lambda)$ over $P$ with multiagent partition
$C^2 = \bigcup_{p\in P} CC_p$ is \temph{distributed}
if:
\begin{enumerate}[partopsep=0pt,topsep=0pt,parsep=0pt]
    \item $C = S^P$ for some set $S$, referred to as \temph{local states}, namely each configuration $c \in C$ consists of a set of local states in $S$ indexed by $P$, in which case we use $c_p \in S$ to denote the local state of $p \in P$ in configuration $c \in C$, and
    \item Any $p$-transition $c\rightarrow c' \in 
    CC_p$ satisfies that $c'_p \ne c_p$ and $c'_q = c_q$ for every $q\ne p \in P$.
\end{enumerate}  
Else $TS$ is \temph{centralized}.
\end{definition}
Namely, in a distributed transition system a $p$-transition (correct or faulty)  can only change the local state of $p$. In other words, even a faulty agent cannot affect the local states of other agents. As a shorthand, we will omit `multiagent' from distributed multiagent transition systems, and instead of presenting a distributed multiagent transition system over $P$ and $S$ as $TS=(S^P,c_0,T,\lambda)$, we will refer to it as the distributed transition system  $TS=(P,S,c_0,T,\lambda)$.

Next, we modify the longest-chain transition system LC  (Example \ref{example:LC}) to become the distributed  transition system for \LCC, LCC, in which each agent has a chain as its local state. 

\begin{example}[LCC: \LCC]\label{example:LCC}
Given a set of agents $P \subseteq \Pi$ and states $S$,  the \LCC distributed transition system LCC $~= (P,(S\times P)^*,c0,T,\lambda)$, has sequences over $S\times P$ as local states, an empty sequence as the initial local state $c0 = \{\Lambda\}^P$, and as $p$-transitions $T$ every $c \rightarrow c'$ where $c'$ is obtained from $c$ by only extending $c_p$, $c'_p = c_p \cdot x \cdot (s,p)$, $s\in S$, and
$c'_q =c_q$ for every $q \ne p \in P$, provided that either
$c_p$ is a longest sequence in $c$ and $x = \Lambda$, or
$c_p\cdot x = c_q \in c$ for some $q\ne p \in P$ and $c_q$ is a longest sequence in $c$.  The liveness condition $\lambda = \{T_p : p \in P\}$ is the multiagent partition over correct transitions.
\qed\end{example}
Note that the \LCC transition system, while distributed, is synchronous (a notion defined formally below), as an agent's ability  to extend its local chain by a certain element depends on the present local states of other agents.
Next,  we show that LCC can implement SCC, making the longest-chain consensus distributed transition system LCC universal for shared-memory multiagent transition systems.

\begin{proposition}\label{proposition:LCC-implements-SCC}
LCC can implement SCC.
\end{proposition}

We noted informally why we consider  LCC synchronous. Next, we define the notions of synchronous and asynchronous distributed transition systems, prove that LCC is synchronous and investigate an asynchronous distributed transition system and its implementation of the LCC.

\subsection{Synchronous and Asynchronous Distributed Multiagent Transition Systems}\label{subsection:mts-sa}

A partial order $\preceq$ over a set of local states $S$  naturally extends to
configurations $C = S^P$ over $P \subset \Pi$ and $S$ by
$c \preceq c'$ for $c, c' \in C$ if $c_p \preceq c'_p$ for every $p \in P$.

\begin{definition}[Distributed Transition System; Synchronous and Asynchronous]\label{definition:multiagent-sa}
%\begin{definition}[Multiagent Distributed Transition Systems]\label{definition:multiagent}
Given agents $\PP \subseteq \Pi$, local states $S$,  and a distributed transition system $TS=(P,S,c_0,T,\lambda)$, then $TS$ is \temph{asynchronous} wrt a partial order $\preceq$ on $S$ if: 
\begin{enumerate}[partopsep=0pt,topsep=0pt,parsep=0pt]
    \item $TS$ is monotonic wrt  $\preceq$, and
    \item for every $p$-transition $c \xrightarrow{} c' \in T$, $T$ also includes the $p$-transition $d \xrightarrow{} d'$ for every $d, d' \in C$ that satisfy the following \textbf{asynchrony condition}:
$$
 c\preceq d \text{ and } (c_p \rightarrow c'_p) = (d_p \rightarrow d'_p)
$$
\end{enumerate}
If no such partial order on $S$ exists, then $TS$ is \temph{synchronous}.
\end{definition}

With this definition, we note that the distributed longest-chain transition system LCC is not asynchronous wrt the prefix relation, as an enabled transition to extend the local chain can become disabled if some other chain extends and becomes longer.  We argue that this is the case wrt any partial order.

\begin{proposition}\label{proposition:LCC-synchronous}
\LCC is synchronous.
\end{proposition}

Next we devise the \AD transition system AD, and prove its universality by using it to implement the synchronous LCC.

We assume a given \emph{payloads function} $\calX$ that maps each set of agents $P$ to a set of payloads $\calX(P)$.  For example, $\calX$ could map $P$ to all strings signed by members of $P$; or to all messages sent among members of $P$, signed by the sender and encrypted by the public key of the recipient; or to all financial transactions among members of $P$. Remember that here $P$ are not `miners' serving transactions by other agents, but are the full set of agents participating the in protocol.

We assume a given \emph{payloads function} $\calX$ that maps each set of agents $P$ to a set of payloads $\calX(P)$.  For example, $\calX$ could map $P$ to all strings signed by members of $P$; or to all messages sent among members of $P$, signed by the sender and encrypted by the public key of the recipient; or to all financial transactions among members of $P$. 

\begin{definition}[Block, $SB$, $\prec_{SB}$]
  A \temph{block} over $P$ is a triple $(p,i,x) \in P \times \calN \times \calX(P)$.  Such a block is referred to as an \temph{$i$-indexed $p$-block with payload $x$}.  The local states function $SB$ maps $P$ to the set of all sets of blocks over $P$ and $\calX(P)$.
    The partial order $\prec_{SB}$ is defined  by $c \preceq_{SB} c'$ if
   $c, c'$ are configurations over $P$ and $SB$ and $c_p \subseteq c'_p$ for every $p \in P$.
\end{definition}
Note that  $c \preceq_{SB} c'$  implies that $c\prec_{SB} c'$ if $c_p \subset c'_p$ for some $p \in P$ and $c_q = c'_q$ for every $q \ne p \in P$.

\begin{example}[AD: \AD]\label{example:AD}
Given a set of agents $P \subseteq \Pi$ and states $S$ that do not include the undefined element $\bot \notin S$,  the \AD transition system, AD $~= (P,B,c0,T,\lambda)$, has local states $B$ being all finite sets of blocks over $P$ and $S\cup \{\bot\}$, an empty set as the initial local state $c0 = \{\emptyset\}^P$, and $T$ has every $p$-transition $c \rightarrow c'$ for every $p \in P$, where $c'$ is obtained from $c$ by adding a block $b=(p',i,s)$ to $c_p$, $c'_p = c_p \cup \{b\}$, 
$p' \in P$, $i \in \calN$, $s\in S\cup \{\bot\}$, and either:
\begin{enumerate}[partopsep=0pt,topsep=0pt,parsep=0pt]
    \item \textbf{$p$-Creates}: $p'=p$, $i=i'+1$, where $i' := \text{ max } \{ j : (p,j,s) \in c_p\}$, or
    \item \textbf{$p$-Receives-$b$}: $p'\ne p$, $(p',i,s) \in c_q \setminus c_p$ for some $q\ne p \in P$.  
\end{enumerate}
The liveness condition $\lambda$ places  transitions with the same label in the same set.
\qed\end{example}
In other words, every agent $p$ can either add a consecutively-indexed $p$-block to its local state, possibly with $\bot$ as payload, or obtain a block it does not have from some other agent.
Note that \AD is asynchronous.
The liveness condition ensures that every correct agent will receive any block created by a correct agent; but it leaves agents the freedom as to which blocks to create.

Next, we explore some properties of \AD: Fault-resilient dissemination and equivocation detection. 
We use `$p$ knows $b$' in a run $r$ to mean that $b \in c_p$ for some $c\in r$.

While in AD agents do not explicitly disseminate blocks they know to other agents, only receive blocks that they do not know from other agents, faulty agents may cause partial dissemination by deleting a block from their local state after only some of the agents have received it. The following proposition states that faulty agents cannot prevent correct agents from eventually sharing all the blocks that they know, including blocks created and partially disseminated by faulty agents.  
\begin{proposition}[AD Block Liveness]\label{proposition:AD-block-liveness}
In an AD run, if a correct agent knows a block $b$ then eventually all correct agents know $b$.
\end{proposition}

\begin{definition}[Equivocation]
An \temph{equivocation} by agent $p$ consists of two $p$-blocks $b=(p,i,s)$, $b'=(p,i',s')$ where $i=i'$ but $s\ne s'$.  An agent $p$ is an \temph{equivocator} in $B$ if $B$ includes an equivocation by $p$.
A set of blocks $B$ is \temph{equivocation-free} if it does not include an equivocation.
\end{definition}

The following corollary states that if an agent $p$ tries to mislead (e.g. double spend) correct agents by disseminating to different agents equivocating blocks, then eventually all correct agents will know that $p$ is an equivocator.
\begin{corollary}[AD Equivocation Detection]
In an AD run, if two blocks $b, b'$ of an equivocation by agent $p$ are each known by a different correct agent,
then eventually all correct agents know that $p$ is an equivocator.
\end{corollary}

Next, we prove that \AD can implement the synchronous distributed longest-chain transition system LCC.  In fact, this implementation offers a naive distributed asynchronous ordering consensus protocol.  Its lack of resilience to equivocation and to fail-stop agents, implied by the FLP theorem~\cite{fischer1985impossibility}, is discussed in the next section. This limitation reflects on the implementation presented here and and not on AD:  The Cordial Miners family of protocols~\cite{keidar2022cordial} employs a more concrete and practical (blocklace-based~\cite{shapiro2022blocklace}) variant of \AD to construct Byzantine fault-resilient order consensus protocols (aka Byzantine Atomic Broadcast) for the models of asynchrony and eventual synchrony.

\begin{proposition}\label{proposition:AD-can-implement-LCC}
AD can implement LCC.
\end{proposition}

The implementation presented is sufficient for the proof but it is naive and not fault resilient.  It is round-based, where all agents participate in every round.  In each round every agent produces a block, with a payload if the agent has any, else without. A round is complete once all agents contributed their blocks. Payloads are ordered according to their round, and the payloads in each round (if any) are ordered lexicographically by agent identifier.
Efficient and fault-resilient implementations~\cite{keidar2022cordial} can also be employed within this framework.

\section{Safety Faults,  Liveness Faults,  and their Resilience}\label{subsection:mtsf}

A safety fault is a subset (or all) of the incorrect transitions, and a liveness fault is a subset of the liveness condition. A computation performs a safety fault $F$ if it includes an $F$ transition. It performs a liveness fault $\lambda' \subseteq \lambda$ if it is not live wrt a set $L \in \lambda'$.
Formally:
\begin{definition}[Safety and Liveness Faults]\label{definition:ts-slf}
Given a transition system $TS=(S,s_0,T,\lambda)$, a \temph{safety fault} is a set of incorrect transitions $F\subseteq S^2 \setminus T$.
A computation \temph{performs a safety fault $F$} if it includes a transition from $F$.  A \temph{liveness fault} is a a subset $\lambda' \subseteq \lambda$ of the liveness condition $\lambda$.
An infinite run \temph{performs a liveness fault $\lambda'$} if it is not live wrt $L$ for some $L \in \lambda'$.
\end{definition}

Note that any safety fault can be modelled with the notion thus defined, by enlarging $S$ and thus expanding the set of available incorrect transitions $S^2$.  Similarly, any liveness fault can be modeled by revising $\lambda$ accordingly.

\begin{definition}[Safety-Fault Resilience]\label{definition:resilience}
Given transition systems $TS=(S,s_0,T\lambda)$, $TS'=(S',s_0',T',\lambda')$ and a safety fault $F \subseteq S'^2 \setminus T'$,
%an agent $p \in \PP$ is \temph{faulty} in a run $r$ of $TS'$ if  $r$ has a faulty $p$-transition, else \temph{correct}. 
a correct implementation $\sigma: S' \xrightarrow{} S$ is \temph{$F$-resilient} if for any live $TS'$ run $r' \subseteq T \cup F$,  the run $\sigma(r')$ is correct.
\end{definition}
In other words, a safety-fault-resilient implementation does not produce incorrect transitions of the specification even if the implementation performs safety faults, and it produces a live run if the implementation run is live.

 Next we compare the resilience of  single-chain consensus SCC and longest-chain consensus LCC to the safety fault in which an agent trashes the chain by adding junk to it. We show that the implementation of the generic shared-memory GS by LCC is more resilient to such faults than the implementation by SCC:  In SCC such a faulty transition terminates the run, violating liveness; in LCC it does not, as long as there is at least one non-faulty agent.

The following Theorem addresses the composition of safety-fault-resilient implementations. See Figure \ref{figure:safety-resilience}.
\begin{theorem}[Composing Safety-Fault-Resilient Implementations]\label{theorem:composing-safety-resilience}
Assume transition systems $TS1=(S1,s1,T1,\lambda1)$,  $TS2=(S2,s2,T2,\lambda2)$, $TS3=(S3,s3,T3,\lambda3)$, correct implementations $\sigma_{21} : S2 \mapsto  S1$ and $\sigma_{32} : S3 \rightarrow S2$, and let $\sigma_{31} := \sigma_{21} \circ \sigma_{32}$. Then: 
\begin{enumerate}[partopsep=0pt,topsep=0pt,parsep=0pt]
    \item If $\sigma_{32}$ is resilient to $F3 \subseteq S3^2 \setminus T3$, then 
    $\sigma_{31}$ is resilient to $F3$.
    \item If $\sigma_{21}$ is resilient to $F2 \subseteq S2^2 \setminus T2$,
    and $F3 \subseteq S3^2 \setminus T3$ satisfies $\sigma_{32}(F3) \subseteq F2$, then
    $\sigma_{31}$ is resilient to $F3$.
    \item These two types of safety-fault resilience can be combined for greater resilience: If $\sigma_{21}$ is $F2$-resilient, $\sigma_{32}$ is $F3$-resilient,  $F3' \subseteq S3^2 \setminus T3$, and $\sigma_{32}(F3') \subseteq F2$, then $\sigma_{31}$ is resilient to $F3 \cup F3'$.
\end{enumerate}
\end{theorem}

\begin{example}[Resilience to Safety Faults in Implementations by SCC and LCC]
For SCC, consider the safety fault $F1$ to be the faulty $q$-transitions $c \rightarrow c \cdot 0$ for every configuration $c$ and some $q\in P$.  For LCC,  consider the safety fault $F2$ to be the faulty $q$-transitions $c_q \rightarrow c_q \cdot 0$ for every configuration $c$ and some agent $q \in P$.
Then a faulty SCC run $r$ with an $F1$ transition cannot be continued, and hence $\sigma_2(r)$ is not live and hence incorrect.  On the other hand, in a faulty LCC run $r$ with $F2$ transitions, the faulty transitions are mapped by $\sigma_{2m}$ to stutter, the run can continue and the implementation is live as long as at least one agent is not faulty.  Note that this holds for the implementation of SCC by LCC, as well as for the composed implementation of GS by LCC, as stated by the following Theorem \ref{theorem:composing-safety-resilience} (the $F3$ case).
\qed\end{example}

Next we consider the implementation of longest-chain consensus LCC by asynchronous block dissemination AD, and it non-resilience to the safety fault of equivocation.

\begin{example}[Non-Resilience to Equivocation of the implementation of LCC by AD]
Consider the implementation $\sigma_3$ of LCC $~= (P,(S\times P)^*,c0,T)$ by AD $~= (P,B,c0,T,\lambda)$, and let $F \subset (B^P)^2$ include equivocations by a certain
agent $p \in P$ for every configuration, namely for every configuration $c \in B^P$ in which $c_p$ includes a $p$-block $b=(p,i,s)$, $F$ includes the $p$-transition $c_p \rightarrow c_p \cup \{b'\}$   for $b' = (p,i,s')$ for some $s' \ne s \in S$. A run $r$ with such an equivocating transition by $p$ may include subsequently a $q$-Receives-$b$ and $q'$-Receives-$b'$ transitions, following which, say in configuration $c'$, the chain computed by $\sigma_3(c')$ for $q$ and for $q'$ would not be consistent, indicating  $\sigma_3(r)$ to be faulty (not safe).
\qed\end{example}

\begin{definition}[Can Implement with Safety-Fault Resilience]
Given transition systems $TS=(S,s_0,T,\lambda)$, $TS'=(S',s_0',T',\lambda')$ and  $F \subseteq S'^2 \setminus T'$,  $TS'$ \temph{can implement $TS$ with $F$-resilience}  if there is an instance $TS''=(S'',s_0',T'',\lambda'') \subseteq TS'$, $F \subset S''\times S''$, and an $F$-resilient implementation $\sigma: S'' \rightarrow S$ of $TS$ by $TS''$.
\end{definition}
The requirement $F \subset S''\times S''$ ensures that the subset $TS''$ does not simply `define away' the faulty transitions $F$.

\begin{definition}[Can Implement with Liveness-Fault Resilience]
Given transition systems $TS=(S,s_0,T,\lambda)$, $TS'=(S',s_0',T',\lambda')$, then $TS'$ \temph{can implement $TS$ with $\bar{\lambda}$-resilience},  $\bar{\lambda}\subseteq \lambda'$,  if there is an instance $TS''=(S'',s_0',T'',\lambda'') \subseteq TS'$,  and an implementation $\sigma: S'' \rightarrow S$ of $TS$ by $TS''$, resilient to $\bar{\lambda}$ restricted to $\lambda''$.
\end{definition}

As an example of resilience to a liveness fault, consider the following:
\begin{example}[Resilience to fail-stop agents of the implementation of SCC and GC by LCC]
Consider LCC $~= (P,(S\times P)^*,c0,T,\lambda)$, and recall that the liveness condition $\lambda = \{T_p : p \in P\}$ is the multiagent partition over correct transitions.
An LCC run $r$  with a liveness fault $\lambda'$ may have all agents $p$ for which $T_p \in \lambda'$ fail-stop after some prefix of $r$.  Still, at least one live agent remains by  the assumption that $\lambda'$ is a strict subset of $\lambda$, and hence $\sigma_2(r)$ is a live (and hence correct) LCC run.  Thus $\sigma_3$ is resilient to any liveness fault of LCC provided at least one agent remains live.  Next, consider the implementation of GC by LCC.  First, we defined an instance SCC1 of SCC to implement GC.  Then we defined LCC1 an instance of LCC to implement SCC1.  Such a composed implementation is resilient to fail-stop agents ($\bar{\lambda}$ in the example above), where their transitions are restricted to LCC1 ($\lambda''$ in the definition above).
\qed\end{example}

\section{Protocols}\label{subsection:protocol} 

Above, we have used the notion of a protocol informally.  Here we make this notion precise.

\begin{definition}[Protocol]\label{definition:protocol}
A \temph{protocol} $\calF$ is a family of multiagent transition systems that has one transition system $TS(P) \in \calF$ over $P$ for every
 $\emptyset \subset P \subseteq \Pi$.  
\end{definition}
Namely, a protocol maps each nonempty set of agents $P \subseteq \Pi$ to a multiagent transition system that specifies the possible correct behaviors of $P$.

In particular, we are interested in distributed protocols in which the set of local states of a transition system over $P$ is a function of $P$.  For example, the local states could be
sequences of messages among members of $P$, or blockchains created and signed by members of $P$, or sets of posts/tweets and threads of responses to them by members of $P$.  Each such set of possible states comes equipped with a partial order that has a minimal element, for example prefix as the partial order for sequences and the empty sequence as the initial state, and subset as the partial order for sets and the empty set as the initial state.

\begin{definition}[Local States Function, $\prec$, Initial State]
A \temph{local states function} $S$ maps every set of agents $P \subseteq \Pi$ to a set of all possible \temph{local states} $S(P)$ $P$. A local states function $S$ has an associated partial order $\prec_{S}$ over its range that is unbounded over $S(P)$ for every $\emptyset \subset P \subseteq \Pi$ and has a minimal element $s0$, referred to as the \temph{initial local state} of $S$.
\end{definition}
Such a local states function $S$ defines for each set of agents $P$ the set of all possible configurations over $P$, as well as the initial configuration over $P$, as follows.

\begin{definition}[Configuration over Local States Function]
Given a  local states function $S$ with partial order $\prec_S$ and minimal element $s0$, given a finite set of agents $P\subseteq \Pi$, a \temph{configuration} $c$ over $P$ and $S$ is a member of $S(P)^P$, namely $c$ consists of a set of local states in $S(P)$ indexed by $P$, with $\{s0\}^P$ being its \temph{initial configuration}.
The partial order  $\prec_S$ on local states induces a partial order on configurations, defined by $c \preceq_S c'$ if $c_p \preceq_S c'_p$ for every $p \in P$.  
\end{definition}

\begin{definition}[Distributed Protocol]\label{definition:distributed-protocol}
Given a local states function $S$ with a partial order $\prec_S$ and minimal element $s0$, a 
\temph{distributed transition system over $P$ and $S$} has configurations over $P$ and $S$ and initial configuration $\{s0\}^P$.
A \temph{distributed protocol $\calF$ over $S$} is a protocol that has a distributed transition system 
$TS(P)=(S(P)^P,\{s0\}^P,T(P),\lambda(P)) \in \calF$
 over $P$ and $S$ for every $\emptyset \subset P \subseteq \Pi$, abbreviated, for a given $P$, as $TS =(P,S,T,\lambda)$.
%
%The \temph{states} of $\calF$ are $S(\calF) := \bigcup_{P \subseteq \Pi}S(P)$.The protocol $\calF$ is \temph{asynchronous} if every transition system in $\calF$ is asynchronous, and is a \temph{subset} of protocol $\calF'$ if for every $P\subseteq \Pi$, $TS(P)$ is a subset of $TS'(P)$.
\end{definition}

Each of the distributed transition systems presented above, GS (Ex. \ref{example:GS}), SCC (Ex. \ref{example:SC}), LCC (Ex. \ref{example:LCC}) and AD (Ex.\ref{example:AD}) can be viewed as members over $P$ of the corresponding protocols $\calG\calS$, $\calS\calC\calC$, $\calL\calC\calC$, and $\calA\calD$.
We illustrate this with the All-to-All Block Dissemination protocol $\calA\calD$.

\begin{definition}[$\calA\calD$:  \AD]\label{definition:AD}
The \temph{All-to-all dissemination protocol} $\calA\calD$ is a protocol over $SB$ that for each $P \subseteq \Pi$ has the transition system $AD= (SB(P)^P,\{\emptyset\}^P,T,\lambda)$, with correct transitions $T$ having a $p$-transition $c \rightarrow c' \in T_p$, $c'_p = c_p \cup \{b \}$, $b= (p',i,x)$, for every   $p, p' \in P$,  $i \in \NN$, $x\in \calX(P)$, and either:
\begin{enumerate}[partopsep=0pt,topsep=0pt,parsep=0pt]
    \item \textbf{Create}: $p'=p$, $i = \text{ max } \{ j : (p,j,x) \in c_p\}+1$, or
    \item \textbf{$q$-Sent-$b$}: $p'\ne p$, $b \in c_q \setminus c_p$ for some $q \in P$,
    provided $i=1$ or $c_p$ has an $(i-1)$-indexed simple $p'$-block.  
\end{enumerate}
The liveness condition $\lambda$ places all $p$-transitions with the same label in the same set, for every $p \in P$.
\end{definition}

\begin{definition}[$\prec_{\calA\calD}$]\label{definition:prec-AD}
A configuration $c$ over $P$ and $SB(P)$, $P\subseteq \Pi$, is \temph{consistent} if for every $p$-block $b \in c$, $b \in c_p$, and it is \emph{complete} if for every $i$-indexed $q$-block $b\in c_p$, $c_p$ includes every $i'$-indexed $q$-block $b'$, for every $1\le i' < 1$, $p,q \in P$.   The partial order $\prec_{\calA\calD}$ is defined  by $c \prec_{\calA\calD} c'$ if
  $c\prec_{SB}c'$ and $c$ and $c'$ are consistent and complete configurations over $P \subseteq \Pi$ and $SB$.
\end{definition}

The notion of a protocol is useful when discussing the relations between different members of the protocol family, for example in defining the notion of a grassroots protocol~\cite{shapiro2023grassroots}.

\section{Conclusions}\label{section:conclusions}

Multiagent transition systems come equipped with powerful tools for specifying distributed protocols and for proving the correctness and fault-resilience of implementations among them.  The tools are best applied if the transition systems are monotonically-complete wrt a partial order, as is often the case in distributed protocols and algorithms.  Employing this framework in the specification of a grassroots ordering consensus protocol stack has commenced~\cite{shapiro2022blocklace}, with sovereign cryptocurrencies~\cite{shapiro2022sovereign} and an efficient Byzantine atomic broadcast protocol~\cite{keidar2022cordial} as the first applications.

%\section*{Acknowledgements}
%I thank Nimrod Talmon, Ouri Poupko, Oded Naor and Idit Keidar and the anonymous referees for discussions and feedback. Ehud Shapiro is the Incumbent of The Harry Weinrebe Professorial Chair of Computer Science and Biology at the Weizmann Institute.  Part of this work was carried out while I was a visiting scholar at Columbia University.

%% Acknowledgments
%%\end{acks}

\newpage

%% Bibliography
%\bibliography{bibfile}
%\bibliographystyle{plain}
\bibliographystyle{splncs04}
\bibliography{bib1}

\newpage

%% Appendix
\appendix

\section{Proofs}\label{appendix-section:proof}

\begin{proof}[of Observation \ref{observation:final-state}]
Assume by way of contradiction that the live computation $r$ is finite and its last state $s$ is not final. Hence there is a correct transition $t$ enabled on $s$, and $r$ violates both liveness requirements:  First, that $r$ has a nonempty suffix in which no correct transition is enabled, since $t$ is enabled on every nonempty suffix of $r$.  Second,that  every suffix of $r$ includes an correct transition, since the suffix that include only $s$ does not. Hence $r$ is not live. A contradiction.
\qed \end{proof}

\begin{proof}[of Proposition \ref{proposition:sigma-correct}]
We prove the proposition by way of contradiction.
Assume that $\sigma$ is locally safe but not safe.  Hence, there is a computation $r' \subseteq T'$ with an incorrect  transition $t \in \sigma(r') \setminus T$.  Consider a prefix $r''$ of $r'$ for which $t \in \sigma(r'')$.  This prefix violates local safety.  A contradiction.

Assume that $\sigma$ is productive but not live.  Then there is a set of transitions $L \in \lambda$
and a computation $r' \subseteq T'$ for which $\sigma(r')$ is not live wrt $L$.  This means that 
in every nonempty suffix of $\sigma(r')$ $L$ is enabled, and there is a suffix of $\sigma(r')$ that does not include an $L$ transition.  This violates both alternative conditions for $\sigma$ being productive: that $r'$ has a nonempty suffix $r''$ such that $L$ is not enabled in $\sigma(r'')$, and that every suffix $r''$ of $r'$  \emph{activates} $L$.   A contradiction.

Assume that $\sigma$ is locally complete but not complete.  Then there is a run $r \subseteq T$ for which there is no run $r' \subseteq T'$ such that $\sigma(r')=r$.  Then there must be a prefix
$\bar{r} \prec r$ of $r$ for which for no run $r' \subseteq T'$,
$\bar{r} \preceq \sigma(r')$.  Thus $\bar{r}$ violates local completeness, a contradiction.
This completes the proof.
\qed \end{proof}

\begin{proof}[of Proposition \ref{proposition:SC-implements-G}]
Given a generic transition system  G $=(S,s0,TG)$ over $S$, we define an instance SC1 of SC and a mapping $\sigma_1$ from SC1 to G that together implement G.
The transition system SC1 $=(S^+,s0,TSC1)$ has the transition $x \cdot s\rightarrow x \cdot s \cdot s' \in TSC1$
for every $x \in S^*$ and every transition $s \rightarrow s' \in TG$.
The mapping $\sigma_1: S^+ \mapsto S$ takes the last element of its input sequence, namely
$\sigma_1(x\cdot s) := s$.

To prove that $\sigma_1$ is correct we have to show that $\sigma_1$ is:
\begin{enumerate}[partopsep=0pt,topsep=0pt,parsep=0pt]
    \item \temph{Locally Safe}:   $s0 \xrightarrow{*} y \xrightarrow{} y' \subseteq TSC1$ implies that  $s0 \xrightarrow{*} x  \xrightarrow{*} x' \subseteq TG$ for $x = \sigma_0(y)$ and $x' = \sigma_0(y')$ in $S$. 
    
    Let $y = s0 \cdot s_1 \cdot \ldots \cdot s_k$, $y' = y \cdot s_{k+1}$, for $k \ge 1$.  For each transition $s0 
    \cdot  s_1 \cdot \ldots \cdot s_i \rightarrow s0 \cdot
    \cdot s_1 \ldots \cdot s_i \cdot s_{i+1} $, $i \le k$,
    the transition $s_i \rightarrow s_{i+1} \in TG$ by definition of $TSC1$. Hence $s0 \xrightarrow{*} x  \xrightarrow{*} x' \subseteq TG$, satisfying the safety condition.

  \item \temph{Productive}:  $TSC1$ is the only set in the liveness condition, and any $TSC1$ transition from any state of SC1 activates $TG$.
  
  \item \temph{Locally Complete}:   $s0 \xrightarrow{*} x\xrightarrow{} x' \subseteq TG$  implies that there are $y, y' \in S^+$ such that $x= \sigma_1(y)$, $x' = \sigma_1(y')$, and $s0 \xrightarrow{*}y \xrightarrow{*} y' \subseteq TSC1$. 
    
    Let $x = s_k$, $x' = s_{k+1}$, $k \ge 1$, and $s0 \rightarrow s_1 \rightarrow \ldots \rightarrow s_k \rightarrow s_{k+1} \in TG$.
    Then $y =  s0\cdot s_1 \cdot \ldots \cdot s_k$ and $y' = y \cdot s_{k+1}$ satisfy the completeness condition.

\end{enumerate}
This completes the proof.
\qed \end{proof}

\begin{proof}[of Proposition \ref{proposition:impelementation-transitivity}]
Assume transition systems $TS1=(S1,s1,T1,\lambda1)$,  $TS2=(S2,s2,T2,\lambda2)$, $TS3=(S3,s3,T3,\lambda3)$ and implementations $\sigma_{21} : S2 \mapsto  S1$ and $\sigma_{32} : S3 \mapsto S2$, and let $\sigma_{31} := \sigma_{21} \circ \sigma_{32}$.  

Assume that $\sigma_{32}$ and $\sigma_{21}$ are safe.  Let $r \subseteq T3$ be a safe $TS3$ run.
Then $\sigma_{32}(r)$ is a safe $TS2$ run by the safety of $\sigma_{32}$, and hence
$\sigma_{21}(\sigma_{32}(r))$ is a safe run by the safety of $\sigma_{21}$.  Hence $\sigma_{31}$ is safe.

Assume that $\sigma_{32}$ and $\sigma_{21}$ are live.  Let $r \subseteq T3$ be a live $TS3$ run.
Then $\sigma_{32}(r)$ is a live $TS2$ run by the liveness of $\sigma_{32}$, and hence
$\sigma_{21}(\sigma_{32}(r))$ is a live run by the liveness of $\sigma_{21}$.  Hence $\sigma_{31}$ is live.

A safe and live run is correct, hence if $\sigma_{32}$ and $\sigma_{21}$ are correct then so is
$\sigma_{31}$.

Assume that $\sigma_{32}$ and $\sigma_{21}$ are complete.  Let $r1 \subseteq T1$ be a correct $TS1$ run.  
By completeness of $\sigma_{21}$ there is a correct $TS2$ run $r2 \subseteq T2$
such that $\sigma_{21}(r2)=r1$.
By completeness of $\sigma_{32}$ there is a correct $TS3$ run $r3 \subseteq T3$
such that $\sigma_{32}(r3)=r2$.
Hence $\sigma_{31}(r3)=r1$, establishing the completeness of $\sigma_{31}$.

This completes the proof.
\qed \end{proof}

\begin{proof}[of Theorem \ref{theorem:sigma-op}]
According to Proposition \ref{proposition:sigma-correct}, to show that a productive $\sigma$ is correct and complete it is sufficient to show that $\sigma$ is:
\begin{enumerate}[partopsep=0pt,topsep=0pt,parsep=0pt]
    \item \temph{Locally Safe}: $s'_0 \xrightarrow{*} y \xrightarrow{} y' \subseteq T'$ implies that  $s_0 \xrightarrow{*} x  \xrightarrow{*} x' \subseteq T$ for $x = \sigma(y)$ and $x'= \sigma(y')$ in $S$.  

    By monotonicity of $TS'$ it follows that $s'_0 \preceq y \prec' y'$; by the Up condition on $\sigma$, it follows that $s_0 \preceq \sigma(y) \preceq \sigma(y')$; by assumption that $TS$ is monotonically-complete it follows that $s_0 \xrightarrow{*} x \xrightarrow{*} x' \subseteq T$ for $x = \sigma(y)$ and $x'= \sigma(y')$ in $S$.  Hence $\sigma$ is safe.

    \item \temph{Locally Complete}: $s_0 \xrightarrow{*} x\xrightarrow{} x' \subseteq T$ 
    implies $s'_0 \xrightarrow{*} y \xrightarrow{*} \in T'$ for some $y, y' \in S'$ such that $x= \sigma(y)$ and $x' = \sigma(y')$.
    
    Let $s_0 \xrightarrow{*} x\xrightarrow{} x' \subseteq T$. By monotonicity of $TS$, $s_0 \preceq x \preceq x'$;  by the Down condition on $\sigma$, there are $y, y' \in S'$ such that $x = \sigma(y)$, $x' = \sigma(y')$, and $y \preceq y'$; by assumption that $TS'$ is monotonically-complete,  $s'_0 \xrightarrow{*} y \xrightarrow{*} y' \subseteq T'$. Hence $\sigma$ is complete. 
\end{enumerate}
This completes the proof of correctness and completeness of $\sigma$.
\qed \end{proof}

\begin{proof}[of Observation \ref{observation:representatives}]
As $TS'$ is monotonically-complete, it has a computation $\hat{\sigma}(x) \xrightarrow{*} \hat{\sigma}(x') \subseteq T'$ that satisfies the Down condition.
\qed \end{proof}

\begin{observation}\label{observation:SC-MC}
SC is monotonically-complete wrt $\preceq$.
\end{observation}
\begin{proof}[of Observation \ref{observation:SC-MC}]
SC is monotonic wrt $\preceq$ since every transition increases its sequence.  Given two sequences $x, x' \in S^*$ such that $x \prec x'$, let $x' = x \cdot s_1 \cdot \ldots \cdot s_k$, for some $k\ge 1$.  Then $x \xrightarrow{*}x'$ via the sequence of transitions $x \rightarrow x\cdot s_1 \rightarrow \ldots \rightarrow x\cdot s_1 \ldots \cdot s_k$.  Hence SC is monotonically-complete.
\qed \end{proof}

\begin{observation}\label{observation:LC-MC}
LC is monotonically-complete wrt $\preceq$.
\end{observation}
The proof is similar to the proof of Observation \ref{observation:SC-MC}.

\begin{observation}[LC Configurations are Consistent]\label{observation:consistent-configurations}
An $n$-chain configuration $c$ is \temph{consistent} if every two chains in $c$ are consistent.
Let $r$  be a run of LC and $c \in r$ a configuration. Then $c$ is consistent.
\end{observation}
\begin{proof}[of Observation \ref{observation:consistent-configurations}]
The proof is by induction on the index $k$ of a configuration in $r$. All empty sequences of the initial configuration of $r$ are pairwise consistent.  Assume the $k^{th}$ configuration $c$ of $r$ is consistent and consider the next $r$ transition $c\rightarrow c' \in TLC$.  The transition adds an element $s$ to one sequence  $x\in c$ that 
either is a longest sequence, or $x\cdot s$ is consistent with another longer sequence $x' \in c$. As all sequences in $c$ are pairwise consistent by assumption, then they are also consistent with $x\cdot s$ by construction.
Hence all sequences of $c'$ are pairwise consistent and hence $c'$ is consistent.
\qed \end{proof}
Hence the following implementation of SC by LC is well-defined.

\begin{proposition}\label{proposition:sigma2-op}
$\sigma_2$ is order-preserving wrt the prefix relation $\preceq$ over consistent $n$-chain configurations and is productive.
\end{proposition}
\begin{proof}[of Proposition \ref{proposition:sigma2-op}]
To show that $\sigma_2$ is order-preserving it is sufficient to show (Proposition \ref{theorem:sigma-op}) that:
\begin{enumerate}[partopsep=0pt,topsep=0pt,parsep=0pt]
    \item \temph{Up condition:} $y \preceq y'$ for $y, y' \in S1$ implies that $\sigma_2(y) \preceq \sigma_2(y')$ and $y \prec y'$ for $y, y' \in S1$ implies that $\sigma_2(y) \prec \sigma_2(y')$
    \item \temph{Down condition:}  $s_0 \xrightarrow{*}x \in T0$, $ x \preceq x'$ implies that there are $y,y' \in S1$ such that $x= \sigma_2(y)$, $x'= \sigma_2(y')$, $c_0 \xrightarrow{* }y \subseteq T1$ and $y \preceq y'$.
\end{enumerate}
Regarding the Up condition, assume that $y \preceq y'$ are consistent
and that $y'_p$ is the unique longest chain in $y'$.  Then
$\sigma_2(y)=y_p \preceq y'_p = \sigma_2(y')$, and if $y \prec y'$
$\sigma_2(y)=y_p \prec y'_p = \sigma_2(y')$.

Regarding the Down condition, define $y_p :=x$, $y'_p := x'$, and $y_q := y'_q := \Lambda$ for every $q \ne p \in \PP$.
Then $x= y_p = \sigma_2(y)$, $x'= y'_p = \sigma_2(y')$, $c0 \xrightarrow{* }y \subseteq T1$ by the same transitions that lead from $s0$ to $x$,  and $y \preceq y'$ by construction.

To see that $\sigma_2$ is productive, note that every LC transition extends one of the chains in a configuration.  Hence, after a finite number of transitions, the next LC chain will extend the longest chain in the configuration, and activate SC.
\qed \end{proof}

\begin{figure}[ht]
\centering
\includegraphics[width=9cm]{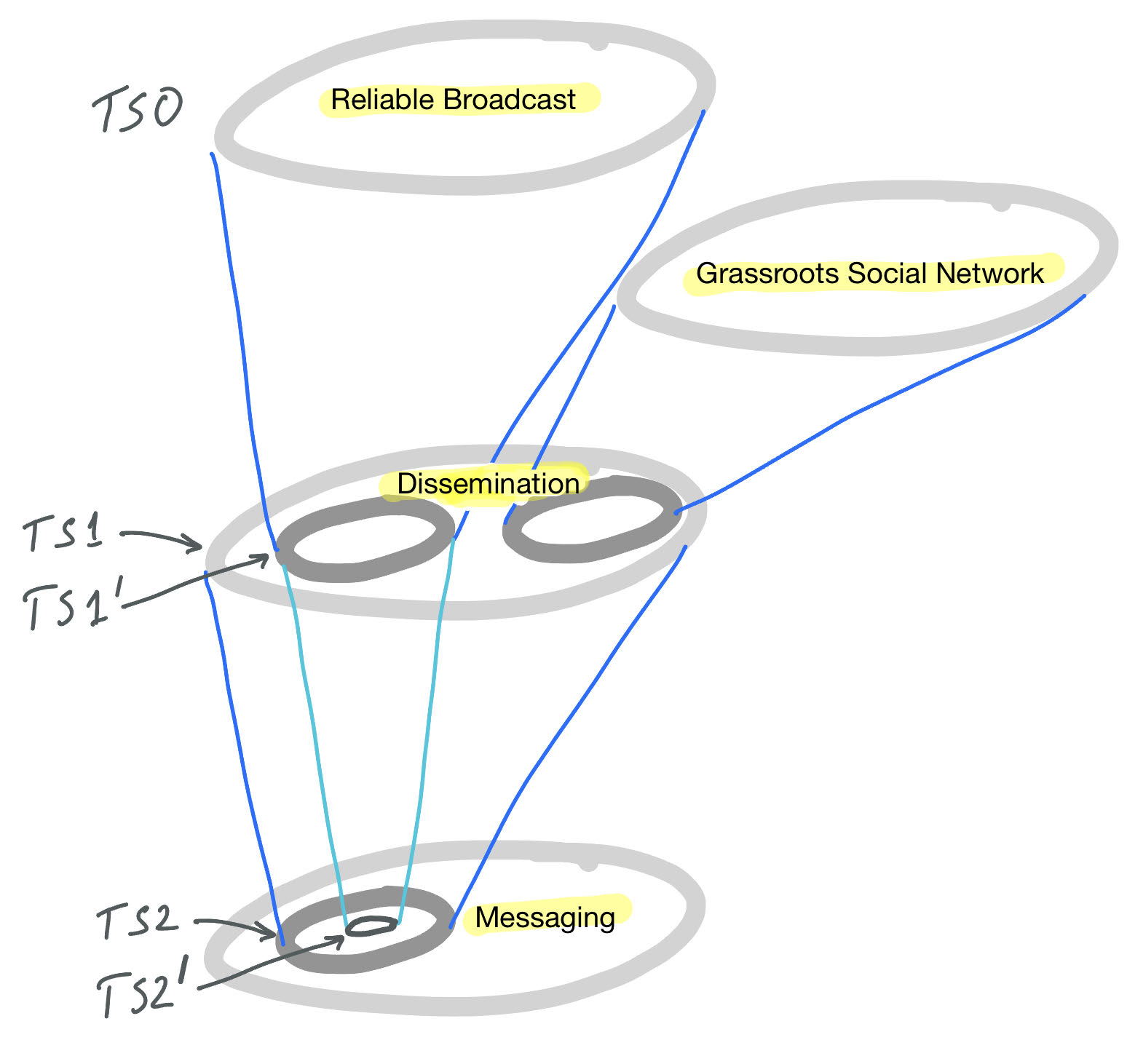}
\caption{Some Steps in the Proof of Proposition \ref{proposition:subset} (with an example in yellow): While $TS2$ (an instance of Messaging) implements $TS1$ (Dissemination), which in turn implements $TS0$ (Reliable Broadcast), $TS1'$ (an instance of Dissemination) is sufficient to implement $TS0$. Hence, it may be more efficient to employ the subset $TS2'$ (of Messaging) instead of the full $TS2$ for the composed implementation of $TS0$. Still, $TS1$ (Dissemination) may have other applications (e.g. grassroots social network,  sovereign cryptocurrencies~\cite{shapiro2022sovereign}), hence it would be useful to implement the entire $TS1$, but then use only the subset $TS2'$ of $TS2$ in the composed implementation of $TS0$. Proposition  \ref{proposition:subset} provides conditions that enable that.}
\label{figure:subset}
\end{figure}

\begin{proof}[of Proposition \ref{proposition:subset}]
Assume $TS1$, $TS2$, $TS1'$, $TS2'$ and $\sigma$ as in the Proposition and that $y \xrightarrow{} y' \subseteq T2 ~\&~ \sigma(y) \in S1'$ implies that $\sigma(y') \in S1'$.  Define $\sigma':C2' \xrightarrow{} S1'$ to be the restriction of $\sigma$ to $C2'$.  We have to show that $\sigma'$ is correct.  To do that, it is sufficient to show that $\sigma'$ is: 
\begin{enumerate}[partopsep=0pt,topsep=0pt,parsep=0pt]
    \item \temph{Locally Safe}:   $s2 \xrightarrow{*} y \xrightarrow{} y' \subseteq T2'$ implies that  $s1 \xrightarrow{*} x  \xrightarrow{*} x' \subseteq T1'$ for $x = \sigma'(y)$ and $x' = \sigma'(y')$ in $S1$. 
    
    This follows from the safety of $\sigma$,  $S1' \subseteq S1$ and the assumption that $y \xrightarrow{} y' \subseteq T2' ~\&~ \sigma(y) \in S1'$ implies that $\sigma(y') \in S1'$.

  \item \temph{Productive}: if any suffix of any infinite correct computation of $TS2'$ 
     activates $T1'$.

    By monotonicity of $TS2'$, any
   infinite correct computation $r$ of $T2'$ from $x'_1$ has a transition $t$ that is strictly increasing, and hence by $\sigma$ satisfying the Up condition, the transition $t$ activates $T1'$.

  \item \temph{Locally Complete}:   $s1 \xrightarrow{*} x\xrightarrow{} x' \subseteq T1'$,  implies that there are $y, y' \in C2'$ such that $x= \sigma'(y)$, $x' = \sigma'(y')$, and $s2 \xrightarrow{*}y \xrightarrow{*} y' \subseteq T2'$. 
    
    By completeness of $\sigma$, there are $y, y' \in C2$ such that $x= \sigma(y)$, $x'= \sigma(y')$, and $s2 \xrightarrow{*} y \xrightarrow{*} y' \subseteq T2$.  By definition of $C2'$ as the domain of $\sigma$, $y, y' \in C2'$.  As $y \xrightarrow{*} y' \subseteq T2$, then $y \preceq_2 y'$.  By assumption that $TS2'$ is monotonically-complete, there is a computation $s2 \xrightarrow{*} y \xrightarrow{*} y' \subseteq T2'$. 
\end{enumerate}
This completes the proof.
\qed \end{proof}

\begin{proof}[of Corollary \ref{corollary:LC-universal}]
Given a generic transition system G over $S$, a correct implementation $\sigma_1$ of G by SC exists according to Proposition \ref{proposition:SC-implements-G}.  The implementation $\sigma_2$ of SC by LC is correct according to Proposition \ref{proposition:LC-implements-SC}.   Then,   Propositions \ref{proposition:impelementation-transitivity} and \ref{proposition:subset} ensure that even though an instance SC1  of SC was used in implementing G, the result of the composition  $\sigma_{21}:= \sigma_2 \circ \sigma_1$ is a correct implementation of G by LC.
\qed \end{proof}

\begin{proof}[outline of Proposition \ref{proposition:SCC-implements-GS}]
The proof is similar to that of Proposition \ref{proposition:SC-implements-G}.
Given a generic shared-memory multiagent transition system  GS $=((S\times \calN)^P,c0,TGS)$ over $P$ and $S$, we define an instance SCC1 of SCC and a mapping $\sigma_{1m}$ from SCC to GS that together implement GS.
The transition system SCC1 $=((S\times P)^+,s0,TSCC1)$ has the $p$-transition $x \cdot (q,s)\rightarrow x \cdot (q,s) \cdot (p,s') \in TSCC1$
for every $x \in (S\times P)^*$, $q\in P$, and every $p$-transition $(s,i) \rightarrow (s',i') \in TGS$.
The mapping $\sigma_{1m}: (S\times P)^+ \mapsto S\times \calN^P$ takes the last element of its input sequence and computes the `program counter' of every agent based on the number of elements by that agent in the input sequence, namely
$\sigma_{1m}(x\cdot (s,p)) := (s,i)$, where $i \in \calN^P$ is defined by $i_q$ being the number of occurrences of $q$ in $x\cdot (s,p)$ for every $q \in P$.
The proof that $\sigma_{1m}$ is correct and complete has the same structure as the proof of $\sigma_1$ in Proposition \ref{proposition:SC-implements-G}.
\qed
\end{proof}

\begin{proof}[outline of Proposition \ref{proposition:LCC-implements-SCC}]
The proof is similar to that of Proposition \ref{proposition:LC-implements-SC}.
We observe that, similarly to SC and LC, both SCC and LCC are monotonically-complete wrt the prefix relation.  For the implementation of SCC by LCC, $\sigma_{2m}$ is the same as $\sigma_2$, except that it returns the longest proper chain in its input, namely a sequence over $S\times P$ (this will prove useful later in showing that $\sigma_{2m}$ is resilient to certain faults).
The proof that $\sigma_{2m}$ is order-preserving wrt $\preceq$ and productive is the same as that of Proposition \ref{proposition:sigma2-op}.
Hence, according to Theorem \ref{theorem:sigma-op}, $\sigma_{2m}$  is correct and complete, which completes the proof.
\qed \end{proof}

\begin{proof}[of Proposition \ref{proposition:LCC-synchronous}]
We have to show that there is no partial order wrt LCC is asynchronous. By way of contradiction, assume that for LCC $~= (P,(S\times P)^*,c0,T)$ there is a partial order $\preceq$ on $(S\times P)^*$ wrt which LCC is asynchronous.  In such a case, by definition,
LCC is monotonic wrt $\preceq$. Let $c$ be a configuration in which $c_p$ is a longest chain,
and let $c\xrightarrow{}\bar{c}$ be a $q$-transition that increases the chain of $q$ so that $\bar{c}_q$ is longer than $c_p$, $q \ne q \in P$.  By monotonicity of LCC, $c \preceq \bar{c}$.  Let $c\rightarrow c'$ be the $p$-transition
$c_p \rightarrow c_p \cdot (s,p)$, with $s \in S$.  Let $d$ be the configuration identical to $c$ except that $d_q := \bar{c}_q$, and let $d'$ be identical to $d$ except that $d'_p := c'_p$.
Hence $d, d'$ satisfy the asynchrony condition (Definition \ref{definition:multiagent-sa}) wrt $c, c'$, and by assumption that LCC is asynchronous wrt $\preceq$ it follows that the $p$-transition $d \rightarrow d' \in T$.  However, this $p$-transition extends $d_p$ in a way inconsistent
with $d_q$, and hence is incorrect.  A contradiction.
\qed \end{proof}

\begin{proof}[of Proposition \ref{proposition:AD-block-liveness}]
If in configuration $c$ there is a block $b$ known by $q$ but not by $p$, both correct, then this holds in every subsequent configuration unless $p$ receives $b$. Hence, due to liveness of $p$-Receives-$b$, either the $p$-Receives-$b$ from $q$ transition is eventually taken, or $p$ receives $b$ through a  $p$-Receives-$b$ transition from another agent. In either case, $p$ eventually receives $b$.
\qed \end{proof}

\begin{proof}[outline of Proposition \ref{proposition:AD-can-implement-LCC}]
Given LCC $~= (P,(S\times P)^*,c0,T)$ and AD $~= (P,B,c0,T,\lambda)$, show that AD and LCC are monotonically-complete wrt $\subseteq$ and $\preceq$ (Propositions \ref{proposition:AD-MC},  \ref{proposition:LCC-MC}), respectively.
Define $\sigma_3$ for each configuration $c \in C$ by $\sigma(c)_p := \sigma'_3(c_p)$, where $\sigma'_3$ is defined as follows.  Given a set of blocks $B$, let $\textit{sort}(B)$ be the sequence obtained by sorting $B$ lexicographically, removing $\bot$ blocks and then possibly truncating the output sequence, where blocks $(p,i,s)$ are sorted first according to the index of the block $i \in \calN$ and then according to the agent $p \in P$, and truncation occurs at the first gap if there is one, namely at the first index $i$ for which the next agent in order is $p$ but there is no block $(p,i,s) \in B$ for any $s \in S \cup \{\bot\}$.
Proposition \ref{propsition:sigma3-op} argues that $\sigma_3$ is order-preserving, which allows the application of Theorem \ref{theorem:sigma-op} and completes the proof.
\qed \end{proof}

Namely $\sigma_3$ performs for each agent $p$ a `round robin' complete total ordering of the set of block of its local state $c_p$, removing undefined elements along the way, until some next block missing from $c_p$ prevents the completion of the total order.

First, we observe that for every configuration $c \in r$ in an AD run $r$, the sequences in $\sigma_3(c)$ are consistent. Note that if $x \preceq y$ and $x' \preceq y$ then $x$ and $x'$ are consistent.

\begin{observation}[Consistency of $\sigma_3$]\label{observation:sigma3-consistency}
Let $r$ be a correct run of AD.  Then for every configuration $c \in r$, the chains of $\sigma(c)$ are mutually consistent.
\end{observation}
\begin{proof}[of Observation \ref{observation:sigma3-consistency}]
First, note that in a correct run $r$, every configuration $c\in r$ is equivocation free.
Also note that $\sigma'_3$ is monotonic wrt $\subseteq$ and $\preceq$, namely  if $B \subseteq B'$ and both $B, B'$ are equivocation free, then $\sigma'_3(B) \preceq \sigma'_3(B')$.
For a configuration $c \in r$,  $c_p \subseteq B(c)$ for every $p \in P$ and hence
$\sigma'_3(c_p) \preceq \sigma'_3(B(c)$, and therefore every two sequences $\sigma'_3(c_p)$, $\sigma'_3(c_q)$ are consistent.
\qed \end{proof}

\begin{proposition}\label{proposition:AD-MC}
 AD is monotonically-complete wrt $\subseteq$.
\end{proposition}
\begin{proof}
 TBC.   
\end{proof}

\begin{proposition}\label{proposition:LCC-MC}
 LCC is monotonically-complete wrt  $\preceq$.
\end{proposition}
\begin{proof}
TBC.   
\end{proof}

Next, we show that $\sigma_3$ is order preserving.
\begin{proposition}\label{propsition:sigma3-op}
$\sigma_3$ is order preserving wrt $\subseteq$ and $\preceq$.
\end{proposition}
\begin{proof}[of Proposition \ref{propsition:sigma3-op}]
According to definition \ref{definition:op-implementation}, we have to prove two conditions.
For the Up condition, we it is easy to see from the definition of $\sigma_3$ that  $c'_1 \subseteq c'_2$ for $c'_1, c'_2 \in B^P$ implies that $\sigma(c'_1) \preceq \sigma(c'_2)$, as the output sequence of the sort procedure can only increase if its input set increases.

For the Down condition, we  construct an AD representative configuration for a LCC configuration so that if the $i^{th}$ element of the LCC longest chain is $(p,s)$, then the AD configuration has the
block $(p,i,s)$, as well as the blocks  $(q,i,\bot)$ for every other agents $q\ne p$.  Specifically,
given a LCC  configuration $c$ with a longest chain $c_l = (s0,p_0),(s_1,p_1),\ldots (s_k,p_k)$ for some $l \in P$, we define the representative AD configuration $c'$ as follows.
First, let $B$ be the following set of blocks $B$:  For each $i \in [k]$
$B$ has the block $(p_i,i,s_i)$ and the blocks $(q,i,\bot)$ for every $q \ne p \in P$.  
Clearly $\sigma'_3(B) = c_l$ by construction.
Let $B^i := \{(p,j,s) \in B : j \le i\}$.  It is easy to see that for each $i\in [k]$,
$\sigma'_3(B^i)$ is the $i$-prefix of the longest chain $c_l$.
Then for each $p \in P$, where $|c_p| = i$, we define $c'_p := B^i$.  Hence, 
$\sigma'_3(c'_p) = c_p$ for every $p\in P$ and thus $\sigma_3(c') = \sigma(c)$.
\qed \end{proof}

\begin{figure}[ht]
\centering
\includegraphics[width=6cm]{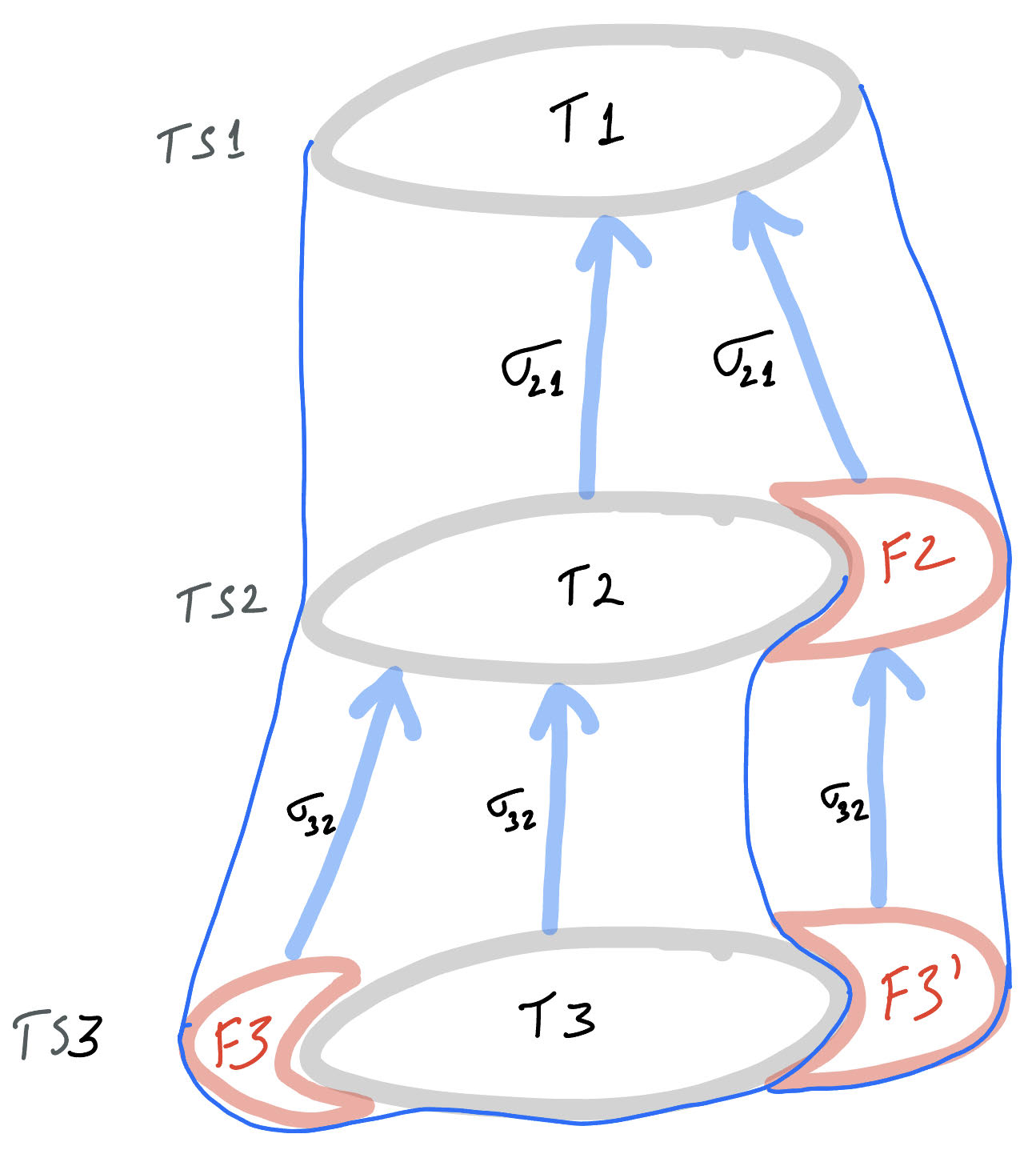}
\caption{Some Steps in the Proof of Theorem \ref{theorem:composing-safety-resilience}:  
$\sigma_{32}$ is resilient to $F3$ and maps $F3'$ to $F2$. $\sigma_{21}$ is resilient to $F2$. As a result, $\sigma_{31}:=\sigma_{21}\circ \sigma_{32}$ is resilient to $F3 \cup F3'$.}
\label{figure:safety-resilience}
\end{figure}
\begin{proof}[of Theorem \ref{theorem:composing-safety-resilience}]
Assume transition systems and implementations as in the theorem statement.
As the composition of live implementations is live, and the assumption is that the runs with safety faults are live, we only argue for safety and conclude correctness.
\begin{enumerate}[partopsep=0pt,topsep=0pt,parsep=0pt]
    \item Assume that $\sigma_{32}$ is resilient to $F3 \subseteq S3^2 \setminus T3$. We argue that $\sigma_{31}$ is resilient to $F3$.  Then  For any $TS3$ run $r \subseteq T3 \cup F3$, the run $\sigma_{31}(r) \in TS1$ is correct, namely $\sigma_{31}(r) \in T1$, since $\sigma_{32}$ is $F3$-resilient by assumption, and hence $r' = \sigma_{32}(r)$ is correct, and $\sigma_{21}$ is correct by assumption, and hence $\sigma_{21}(r)$ is correct, namely $\sigma_{31}(r) = \sigma_{21} \circ \sigma_{32}(r) \in T1$.

     \item Assume $\sigma_{21}$ is resilient to $F2 \subseteq S2^2 \setminus T2$,
    and $F3 \subseteq S3^2 \setminus T3$ satisfies $\sigma_{32}(F3) \subseteq F2$.  We argue that
    $\sigma_{31}$ is resilient to $F3$. For any $TS3$ run $r \subseteq T3 \cup F3$, the run $\sigma_{32}(r) \in TS2 \cup F2$ by assumption.  As $\sigma_{21}$ is $F2$-resilient by assumption, the run $\sigma_{21}\circ\sigma_{32}(r)=\sigma_{31}(r)$ is correct.
    
     \item Assume that $\sigma_{21}$ is $F2$-resilient, $\sigma_{32}$ is $F3$-resilient,  $F3' \subseteq S3^2 \setminus T3$, and $\sigma_{32}(F3') \subseteq F2$. We argue that $\sigma_{31}$ is resilient to $F3 \cup F3'$. For any $TS3$ run $r \subseteq T3 \cup F3 \cup F3'$, 
    the run $\sigma_{32}(r) \in TS2 \cup F2$ by assumption.  As $\sigma_{21}$ is $F2$-resilient by assumption, the run $\sigma_{21}\circ\sigma_{32}(r)=\sigma_{31}(r)$ is correct.
\end{enumerate}
\qed \end{proof}

\begin{proof}[of Proposition \ref{proposition:AD-block-liveness}]
If in configuration $c$ there is a block $b$ known by $q$ but not by $p$, both correct, then this holds in every subsequent configuration unless $p$ receives $b$. Hence, due to liveness of $p$-Receives-$b$, either the $p$-Receives-$b$ from $q$ transition is eventually taken, or $p$ receives $b$ through a  $p$-Receives transition from another agent. In either case, $p$ eventually receives $b$.
\qed \end{proof}

\end{document}